\documentclass{JHEP}
\title{On $(\alpha')^2$ corrections to the D-brane action for non-geodesic
world-volume embeddings}
\input{epsf}
\usepackage{epsfig}
\author{A. Fotopoulos 
\\ Department of Physics
\\ Northeastern University, Boston, MA 02115, U.S.A.
\\ and
\\Centre de Physique Th\'eorique \footnote{Unite mixt\'e du CNRS et de l' 
Ecole Polytechnique, UMR7644}
\\Ecole Polytechnique, 91128 Palaiseau, France\thanks{Current address}.
\\ E-mail: \email{angelos.fotopoulos@cpht.polytechnique.fr}
} \abstract{ In \hepth{9903210}
(curvature)$^2$ terms of the effective D-brane action were derived
to lowest order in the string coupling. Their results 
are correct up to ambiguous terms which involve
the second fundamental form of the D-brane. We compute five point
string amplitudes on the disk. We compare the subleading order in
$\alpha'$ of the string amplitudes with the proposed lagrangian
of \hepth{9903210} supplemented by the ambiguous terms. The
comparison determines the complete form of the gravitational terms
in the effective D-brane action to order ${\cal
O}(\alpha^{'\;2})$. Our results are valid for arbitrary ambient
geometries and world-volume embeddings.} 
\preprint{ \hepth{0104146}\\
CPHT-S021.0401}
\begin{document}
\newcommand{\lp}{\left(}
\newcommand{\rp}{\right)}
\newcommand{\blp}{\biggl(}
\newcommand{\brp}{\biggr)}
\newcommand{\ze}{\zeta}
\newcommand{\apr}{\alpha'}
\newcommand{\al}{\alpha}
\newcommand{\bal}{\bar{\alpha}}
\newcommand{\si}{\sigma}
\newcommand{\de}{\delta}
\newcommand{\bpartial}{ \bar{\partial}}
\newcommand{\bz}{\bar{z}}
\newcommand{\bw}{\bar{w}}
\newcommand{\bh}{\bar{h}}
\newcommand{\bL}{\bar{L}}
\newcommand{\bp}{\bar{p}}
\newcommand{\bpsi}{\bar{\psi}}
\newcommand{\beps}{\bar{\epsilon}}
\newcommand{\grad}{\bigtriangledown}
\newcommand{\cN}{{\cal N}}
%
\section{Introduction}\label{intro}

The $D$-brane low energy effective action for a single brane 
describes the field theory of a gauge boson
$A_{\al}$ and scalar fields $Y^i$, as well as their interactions
with the closed string modes of the bulk \cite{ callan87,LEIGH, 9510017, 9510135} (for reviews on
D-branes see \cite{bachas, TASI}).
For a system of N coincident $D$-branes the 
low energy effective action is generalized to a non-abelian theory with gauge group $U(N)$.
Since we have a supersymmetric theory there is of course a fermionic analog of these 
actions but we will not deal with it in this paper. 

A low energy effective action is obtained in field theory by integrating out
massive modes above some energy scale. This modifies the full action of
the theory by introducing higher derivative terms which encode the
effect of integrating out the massive modes. Effective actions
have a range of validity up to the scale above which we integrate
out. The $D$-brane low energy effective action
is the action of the massless modes. Obviously it receives
higher order corrections. Therefore, for energies much smaller
than the scale of the massive string modes $\frac{1}{\apr}$, the effective action
is given as an expansion in powers of $\apr$ the fundamental energy scale 
in string theory. 

In the case of the D-brane the low energy effective action has two parts. The parity even
part is given by the Dirac-Born-Infeld (DBI) \cite{LeighA,Fradkin} action
and the parity odd by the Wess-Zumino (WZ) \cite{9510017,Green, Cheung, Minasian} action. 
The Wess-Zumino action describes the coupling of $D$-branes to Ramond-Ramond fields.
The form of this action is determined by requiring
cancellation of the chiral gauge anomaly from the field theory on the brane 
against the gravitational anomaly from the bulk for intersecting 
D-branes \cite{Green, Cheung, Minasian}. The WZ term takes the form:
\begin{eqnarray} \label{WZaction}
\int_{M^{(p+1)}} \; {\cal L}_{WZ}^{(p)}= T_{(p)} \int_{M^{(p+1)}}
\; C \wedge tr_{N} \blp e^{2 \pi \apr F} \brp \wedge \blp \frac{
\hat{A}(4 \pi^2 \apr R_T)}{ \hat{A}(4 \pi^2 \apr R_N)}\brp^{1 \over
2}
\end{eqnarray}
where $C= C^{(0)}+C^{(1)}+ \dots C^{(9)}$ are the R-R p-form potentials
pulled-back on the D-brane with odd forms contributing in the IIA
theory and even forms in the IIB theory. The trace of the field
strength $F_{\al \beta}$ is over the U(N) gauge indices for N
coincident D-branes. $\hat{A}$ is the Dirac "roof" genus and its
square root has the expansion:
\begin{eqnarray} \label{roof}
\sqrt{\hat{A}(R)}= 1- \frac{1}{48} p_1(R)+ \frac{1}{2560} p_1^2(R)
- \dots
\end{eqnarray}
Another method of finding these couplings
is by considering scattering of gravitons on the D-branes in the
boundary state formalism \cite{dv9707,anomalous,normal,stefanski}.
   
The DBI action:
\begin{eqnarray}\label{DBI}
S_{DBI}= T_{(p)} \int dx^{p+1} e^{- \phi} \sqrt{ det \blp (G_{\mu
\nu}+ B_{\mu \nu}) \partial_{\alpha} X^{\mu} \partial_{\beta}
X^{\nu}
+ 2 \pi \apr F_{\alpha \beta} \brp} 
\end{eqnarray}
describes the coupling of the brane modes to the NS-NS
sector bulk fields: $\phi, g_{\mu \nu}, B_{\mu \nu}$.
The Born-Infeld action for a single brane in flat background has an
expansion which for $\apr \to 0$ reduces to the $U(1)$ Yang-Mills
theory. For N coincident D-branes the DBI becomes 
a non-abelian field theory whose $\apr \to 0$ limit 
is an SU(N) SYM theory. The expansion includes also terms of the form $(F)^n$  
which give contact interactions of $n$ particle scattering in the
effective field theory. The DBI action should be supplemented
by higher derivative corrections, in the sense that they vanish for 
$F$ constant. In the curved brane case, which is the main 
interest of this paper, derivative corrections involve pull-backs of derivatives 
of the NS-NS background fields and $\partial^n (\partial_{\al} X^{\mu})$ corrections,
where $\partial_{\al}X^{\mu}$ the embedding of the brane in the ambient
spacetime; all of which vanish for constant background and embedding. 
The issue of finding the form of these derivative corrections 
has been addressed in \cite{BachasCurvature,Wyllard}. In \cite{BachasCurvature} 
$\apr^2$ corrections to the DBI for non-constant backgrounds were determined, while in 
\cite{Wyllard} the $\apr^2$ corrections involve derivatives of the gauge field strength
F and of the embedding $\partial X$.     

There are several ways to compute corrections to the D-brane effective action.
One of them involves computing renormalization group 
beta functions for the field theory
of strings on the world-sheet. Consistency conditions (superconformal
invariance) impose that these beta functions should vanish. From these
conditions we can find equations of motion for the background
fields. This method was used to derive the Born-Infeld action
in \cite{LeighA}.

Another way to determine effective actions is by expanding string
amplitudes in powers of the string scale $\apr$  and
looking for terms in the effective action to reproduce this
expansion. We use this method for higher order derivative
corrections since the beta function computation is quite more
involved requiring amplitudes with at least three loops. This
method has also been used to check the validity of the Born-Infeld
action. It has been found through such amplitude computations that the
non-abelian Born-Infeld encounters problems when expanded beyond
the leading order in $\apr$ \cite{DST}. However, in this paper we will focus on
the higher $\apr$ corrections to the single (abelian) $D$-brane action.
In \cite{BachasCurvature} $\apr^2$ corrections to DBI were found by comparison to the 
$\apr$ expansion of the four-point functions computed in \cite{GarScatter}. This way they
arrived to the following langrangian up to order $\apr^2$:
 \begin{eqnarray}
&L^{(p)} =  T_{p} \, e^{-\phi} \sqrt{\tilde{g}} [ 1-\frac{1}{24} \frac{ (4
\pi^2 \alpha ^{'} ) ^{2} }{ 32 \pi ^{2}} ( (R_{T})_{\alpha \beta
\gamma
\delta }(R_{T})^{\alpha \beta \gamma \delta } \nonumber \\
&-2(R_T)_{\alpha \beta }(R_T)^{\alpha \beta }-(R_{N})_{\alpha
\beta i j }(R_{N})^{\alpha \beta i j } +2 \bar{R}_{i j} \bar{R}^{i
j})] \nonumber
\end{eqnarray}
where $R_T$ and $R_N$ are the Riemann tensors constructed from the world-volume and 
normal bundle connections (see Appendix \ref{submanifold}).

The corrections computed in \cite{BachasCurvature} 
were ambiguous because certain combinations of the proposed derivative terms turned out
to vanish for linear expansions in the fields. We will find out that there are five 
ambiguous terms in the D-brane action which cannot be fixed using the constraints 
deduced from the four-point functions computed in \cite{BachasCurvature}. Three of 
these terms are total derivatives for linear expansion of the curvature and second 
fundamental form $\Omega^i_{\al \beta}$ tensors \footnote{see Appendix \ref{submanifold} for definition of 
second fundamental form tensor} and we denote their coefficients by $c_1, \ c_2, \ c_3$. 
The other two ambiguous terms give vanishing contribution to the four-point functions 
because they involve tensors which vanish due to the lowest order equations of motion and we 
denote their coefficients by $c_4, \ c_5$. The ambiguous terms can be summarized by the langrangian:
 \begin{eqnarray}
&L^{ambig}= \displaystyle  [ c_1\blp R_{\alpha \beta \gamma
\delta }R^{\alpha \beta \gamma \delta } -4 \hat{R}_{\alpha \beta
}\hat{R}^{\alpha \beta }+ \hat{R}^2 \brp \nonumber \\
&+c_2( 4 R_{\alpha \beta \gamma \delta} (\Omega^{\alpha \gamma}
\cdot \Omega ^{\beta \delta}) -8 \hat{R}_{\alpha \beta}
(\Omega^{\alpha \gamma} \cdot \Omega ^{\;\beta}_{\gamma})+2
\hat{R}(\Omega^{\alpha
\beta} \cdot \Omega _{\alpha \beta}) ) \nonumber \\
&+ c_3 (2(\Omega_{\alpha \gamma} \cdot \Omega_{\beta \delta})
(\Omega^{\alpha \gamma} \cdot \Omega^{\beta \delta}) -2
(\Omega_{\alpha \gamma} \cdot \Omega_{\beta \delta})
(\Omega^{\alpha \delta} \cdot \Omega^{\beta \gamma}) \nonumber \\
&-4(\Omega_{\alpha \gamma} \cdot \Omega_{\beta}^{\; \gamma})
(\Omega^{\alpha \gamma} \cdot \Omega^{\;\beta}_{\gamma}) +
(\Omega^{\alpha \beta} \cdot \Omega_{\alpha \beta})(\Omega^{\gamma
\delta} \cdot \Omega_{\gamma \delta}) )\nonumber \\
&+ c_4((\Omega_{\gamma}^{\; \gamma} \cdot
\Omega_{ \alpha}^{\; \beta})(\Omega^{\; \alpha}_{\delta} \cdot
\Omega_{\beta}^{\; \delta})) +c_5(\hat{R}_{\alpha \beta}
(\Omega^{\alpha \beta} \cdot \Omega ^{\; \gamma}_{\gamma}))] \nonumber
\end{eqnarray}

In order to determine these ambiguous terms one has to compute higher point amplitudes. 
In this paper we will extend the analysis of \cite{BachasCurvature} to determine the complete
form of the $\apr^2$ corrections to the DBI for curved backgrounds and non-trivial
embeddings. We shall find out that the constraints derived from our five-point functions
computations are sufficient to determine the ambiguous coefficients $c_1, \ c_2, \ c_3$ if we use
in addition a duality argument described in \cite{BachasCurvature}. The remaining coefficients
$c_4, \ c_5$ will remain undetermined since they give vanishing contribution to the amplitudes 
we will consider. Our computations are at disk-level which is the lowest order of the perturbative
expansion of the D-brane action in the string coupling constant. 

In section \ref{preliminaries} we give a short overview of the
basic machinery for computing string amplitudes on the disc. 
In section \ref{1cl3o} we compute the three scalar with one 
graviton string amplitude. In section \ref{normalization} we 
compare the $\apr \to 0$ limit of the string amplitude computed in 
section \ref{1cl3o} with the corresponding field theory expression from
the DBI action in order to fix the normalization. This way we have a check of our
computations as well. In section \ref{corrections1} we present the ambiguous
terms of the D-brane action to order $\apr^2$ and compute the contribution
of the proposed lagrangian to the three scalar with one 
graviton scattering. Consequently we expand the string amplitude 
to subleading order in $\apr$ and compare with the 
field theory amplitude to obtain a relation among the $c_2$ and $c_3$ coefficients. 
We shall also find out that $c_4$ cannot be determined from this amplitude.
We repeat the above analysis for the scattering of
two gravitons with one scalar in sections \ref{2cl1o} and \ref{corrections2}
and find a similar relation among $c_1$ and $c_2$ as well as that $c_5$ cannot be determined
with our methods. Subsequently we use a non-perturbative argument described in \cite{BachasCurvature}
to fix the value of the ambiguities $c_1, \ c_2, \ c_3$. At the end we are left with 
the ambiguities $c_4, \ c_5$ which are proportional to the lowest order equations of motion 
for the scalar fields and in section \ref{conclusion} we comment on this result and find a possible 
explanation of our inability to fix them. We show that such terms
remain arbitrary because they can be eliminated by employing field redefinitions under which
the string S-matrix is invariant \cite{theisen}. The appendices contain integral formulas 
needed for the string amplitudes and a review of the geometrical characteristics of submanifolds 
used in constructing candidate derivative terms for the D-brane action.    
 
\section{Preliminaries}\label{preliminaries}
 In space-time D-branes are represented as
 static p-dimensional defects. As a result of these defects we must impose different
 boundary conditions, on the world-sheet boundary, to coordinates tangent and
 normal to the D-brane.
  \begin{eqnarray} \label{NeuDir}
  \partial_{\perp}X^{\alpha}\mid_{\partial \Sigma}=0 \nonumber \\
  X^{i}\mid_{\partial \Sigma}=0
  \end{eqnarray}
  The lower case Greek indices , ($ \alpha =0,1 \dots ,p $),
  correspond to directions parallel to the brane and the lower Latin ones, ($ i=p+1,
  \dots, 9$) to normal coordinates. These constrains are respectively Neumann
and Dirichlet boundary conditions.

When calculating tree string amplitudes we evaluate the partition
function on a world-sheet with the topology of a disc. Before
presenting details of our calculation, we will review the basic
formalism of string vertex operators and their expectation values
on the disc. We follow closely the review of \cite{hashimoto} and
references therein \cite{GarBI, malda}.

It turns out that it is convenient to introduce the conformally
equivalent description of the disc on the upper half complex plane
for string amplitude computations. We chose this representation of
the disc because it is easier to evaluate string correlators on
the half complex plane. We denote the upper half plane as $
\cal{H}^{+} $ and using radial coordinates $z$ on the half-complex
plane, the real axis becomes the world-sheet boundary.

The string operators for an NS-NS
 massless closed string have the following general form
\begin{eqnarray}\label{GenOper}
V(z, \bar{z})= \epsilon_{\mu \nu} :V_{s}^{\mu}(z): \,
:V_{s}^{\nu}(\bar{z}):
\end{eqnarray}
where $\mu=0,1,\dots,9 \ and \ s=0,-1$ denotes the superghost
charge or equivalently the picture in which the operator is in.
The total superghost charge on the disk is required to be
$Q_{sg}=-2$ as a consequence of the requirement for
superdiffeomorphism invariance. The holomorphic(left moving) parts
are given by
\begin{eqnarray}\label{PicOper}
V_{-1}^{\mu}(p,z)= e^{-\phi(z)} \psi^{\mu}(z) e^{ \imath p \cdot
X(z)}\\
V_{0}^{\mu}(p,z)=( \partial X^{\mu}(z) + \imath p \cdot \psi(z)
\psi^{\mu}(z))e^{ \imath p \cdot X(z)} \nonumber
\end{eqnarray}
and similar expressions for the anti-holomorphic part. The
expectation values of string vertices are found using the
following correlators:
\begin{eqnarray}\label{HolCor}
&\langle X^{\mu}(z) X^{\nu}(w) \rangle= -\eta^{\mu \nu}\log (z-w) \nonumber \\
&\langle \psi^{\mu}(z) \psi^{\nu}(w) \rangle = - \frac{\eta^{\mu
\nu}}{z-w} \\
&\langle \phi(z) \phi(w) \rangle = - \log (z-w) \nonumber
\end{eqnarray}
Because of the boundary conditions we have non-trivial
correlators between right and left moving strings
\begin{eqnarray}\label{HolAntiCor}
&\langle X^{\mu}(z) \bar{X}^{\nu}(\bar{w}) \rangle= -D^{\mu \nu}\log (z-\bar{w})
\nonumber \\
&\langle \psi^{\mu}(z) \bar{\psi}^{\nu}(\bar{w}) \rangle = -
\frac{D^{\mu \nu}}{z-\bar{w}}  \\
&\langle \phi(z) \bar{\phi}(\bar{w}) \rangle = - \log (z-\bar{w})
\nonumber
\end{eqnarray}
where $ D_{\nu}^{\mu}$ is a diagonal matrix with +1 for directions
tangent to the world-volume and -1 for transverse directions. We
raise and lower indices using $ \eta^{\mu \nu} $\cite{GarScatter}.
At this point it is useful to define two more projection matrices
$ V_{\nu}^{\mu}$ and $ N_{\nu}^{\mu} $ which project on the
tangent and normal to the brane spaces respectively. These
projection matrices satisfy the following identities
\begin{eqnarray}
D_{\mu \nu}= V_{\mu \nu} - N_{\mu \nu} \\
\eta_{\mu \nu} = V_{\mu \nu} + N_{\mu \nu} \nonumber
\end{eqnarray}
 By extending the definition of the fields to the whole complex plane
 \cite{hashimoto} we can write all our vertices in terms of left moving string
operators by making the substitutions
\begin{eqnarray} \label{substitute}
\bar{X}^{\mu}(\bar{z}) \rightarrow D^{\mu}_{\nu} X^{\nu}(\bar{z})
\ \ \bar{\psi}^{\mu}(\bar{z}) \rightarrow D^{\mu}_{\nu}
\psi^{\nu}(\bar{z}) \ \ \bar{\phi}(\bar{z}) \rightarrow
\phi(\bar{z}) \nonumber
\end{eqnarray}
where $ z \in \cal{H}^{+}$. This way we use standard correlators (\ref{HolCor}) in string amplitude computations.
 After these replacements we write the vertex operators as follows:
\begin{eqnarray} \label{GravOper}
V(z, \bar{z})= (\epsilon D)_{\mu \nu} :V_{s}^{\mu}(p,z): \,
:V_{s}^{\nu}(Dp,\bar{z}):
\end{eqnarray}
Similar manipulations allow us to write the open string vertex
operators for Neumann and Dirichlet conditions \cite{hashimoto}
\begin{eqnarray} \label{OpenOper}
V_{-1}^{\mu}(2k,z)= e^{-\phi(z)} \psi^{\mu}(z) e^{ \imath 2k \cdot
X(z)}\\
V_{0}^{\mu}(2k,z)=( \partial X^{\mu}(z) + \imath 2k \cdot \psi(z)
\psi^{\mu}(z))e^{ \imath 2k \cdot X(z)} \nonumber
\end{eqnarray}
with z on the real axis and the momentum $k^{\mu} $ is restricted
to be tangent to the brane. The vertex operators (\ref{OpenOper}) are with momenta
$2k$ since we have set $\alpha'=2$  even though they are open strings.
 The general expression of a string amplitude is
\begin{eqnarray} \label{StrAmp}
&A= \displaystyle\int \frac{ d^2 \!z_{J} d^2 \! x_{I}}{V_{CKG}} \langle
\prod_{I=1}^{n} : V(x_{I}): \,\prod_{J=1}^{m} :V(z_{J},
\bar{z}_{J}): \, \rangle
\end{eqnarray}
where n, m are the number of open and closed string operators,
respectively. The diffeomorphisms group CKG (Conformal Killing Group)
 can be used to fix the position of $n_{K} $ operators, where $n_{K} $ the number of
conformal killing vectors of the world-sheet surface. In fixing
these positions we introduce $n_{K} $ fermionic ghosts.

Therefore for the disc with $SL(2,R)$
CKG we have to insert three c-ghosts in the fixed position
operators. Their correlator is given by
\begin{eqnarray}\label{GhostCor}
\langle c(z_1) c(z_2)c(z_3) \rangle = C_{D_2}^{ghost} (z_1 - z_2)
(z_1 - z_3)(z_2 - z_3)
\end{eqnarray}
where $C_{D_2}^{ghost}$ is a normalization constant
\cite{POLCHINSKI}

\section{Graviton-three Scalar scattering amplitude}
\label{1cl3o}
In this section we will calculate the scattering amplitude of one
graviton with 3 world-volume scalars. As explained in the
introduction this amplitude expanded to subleading order in the
momenta will help us determine some of the ambiguity of the
$\alpha ^{'2}$ corrections computed in \cite{BachasCurvature}. We explain
more on these ambiguities in section \ref{corrections1}. The
graviton has momentum in the bulk space-time and the scalar fields
can only propagate parallel to the p-brane.  We insert one closed
string vertex operator in the -1 picture and the open string
vertices in the 0 picture. The amplitude has the following
expression
\begin{eqnarray} \label{3+1Amp}
A \sim \int d^2 \!z \langle \prod_{I=1}^{3} : c(x_{I})
V_{0}^{i}(x_{I}): \, :V_{-1}^{ \mu}(z):\,:V_{-1}^{ \nu}(\bar{z}):
\, \rangle \ze_{I i} (\epsilon D)_{\mu \nu} + ( 2 \leftrightarrow
3)
\end{eqnarray}
where  the index I = 1,2,3 denotes the scalars. The exchange of the
particles 2 and 3 in the last line is necessary since the $SL(2,R)$
 diffeomorphisms group  does not change
the cyclic ordering of the open string vertex operators on the disc
boundary (real axis). For the graviton
amplitude in the abelian case it turns out that this exchange
results in a factor of two in our amplitude. For non-abelian cases
and the Kalb-Ramond field this will have a non-trivial effect on the
amplitude.

We use the following kinematic invariants to describe this
scattering process:
\begin{eqnarray} \label{Kinem}
&s=-4k_{1}k_{2} \qquad t=-4k_{1}k_{3} \qquad u=-4k_{2}k_{3} \nonumber \\
&q^{2}= ( Vp )^2 = -\blp \frac{s+t+u}{2} \brp 
\end{eqnarray}
where $k_1, k_2 , k_3 $ are the three open string momenta and $ p ^{\mu}$
is the momentum of the closed string. To derive the last equation we used the
conservation of momentum on the world-volume
\begin{eqnarray} \label{ConsMom}
\blp k_{1} + k_{2} + k_{3} +  (Vp) \brp  ^{\mu} = 0
\end{eqnarray}
The usual gauge transversality conditions for the graviton and
scalar fields translate to the following conditions on the momenta
and polarizations:
\begin{eqnarray} \label{EqnMot}
\ze _{I} ^{\mu} k_{I \mu} = 0 \qquad \epsilon ^{\mu \nu} p_{\nu} = 0 
\qquad \epsilon_{\mu}^{\mu} = 0 
\end{eqnarray}
 We use the expressions from section \ref{preliminaries}
for the vertex operators to expand the expression appearing inside
the integral in (\ref{3+1Amp}). The integrand in (\ref{3+1Amp})
becomes
\begin{eqnarray} \label{3+1A2}
&\langle \; :c(x_1)(\partial X^i +2i (k_1 \cdot \psi)\psi^i)(x_1)e^{2ik_1 \cdot X(x_1)}: \; :c(x_2)(\partial X^i +2i (k_2
 \cdot \psi)\psi^i)(x_2) \nonumber \\
&e^{2ik_2 \cdot X(x_2)}: \; :c(x_3)(\partial X^i +2i (k_3  \cdot \psi)\psi^i)(x_3)e^{2ik_3 \cdot X(x_3)}: \nonumber \\
&:e^{-\phi(z)} \psi^{\mu}(z) e^{ip \cdot X(z)}: \; :e^{-\phi(\bar{z})} \psi^{\nu}(\bar{z})e^{i(Dp) \cdot X(\bar{z})}: \; \rangle
\ze_{1i} \ze_{2j} \ze_{3l} (\epsilon D)_{\mu \nu}  \nonumber \\
&+ ( 2 \leftrightarrow 3)
\end{eqnarray}
Expanding the terms in the amplitude we get four different types
of path integrals we need to evaluate. We use the expressions from
section \ref{preliminaries} for the correlators to evaluate the
expression appearing inside the integral in (\ref{3+1A2}).
Combining all terms we get
\begin{eqnarray}
&\{ \eta^{\mu \nu} ( - \frac{p^l \eta^{ij}}{(x_1-x_2)^2|x_3-z|^2} +4 \frac{p^i \eta^{jl} k_2 k_3}{(x_2-x_3)^2 |x_1-z|^2})
 2[ -\frac{\eta^{ij}}{(x_1-x_2)^2} \frac{\eta^{l \mu} k_3^{\nu} - \eta^{l \nu} k_3^{\mu}}{|x_3-z|^2} - 4\frac{\eta^{ij}}{x_1-x_2} \blp
-\frac{\eta^{l \mu}}{x_3-z}( \frac{k_3^{\nu} (k_1 k_2)}{(x_1-x_2)(x_3- \bar{z})} \nonumber \\
&-\frac{k_2^{\nu} (k_1 k_3)}{(x_1-x_3)(x_2- \bar{z})} + \frac{k_1^{\nu} (k_3 k_2)}{(x_2-x_3)(x_1- \bar{z})}) + \frac{\eta^{l \nu}}{x_3-
\bar{z}}( \frac{k_3^{\mu} (k_1 k_2)}{(x_1-x_2)(x_3- z)}
-\frac{k_2^{\mu} (k_1 k_3)}{(x_1-x_3)(x_2- z)}+\frac{k_1^{\mu} (k_1 k_3)}{(x_2-x_3)(x_1- z)}) \brp \nonumber \\
&+4 \frac{p^i(z-\bar{z})}{(x_2-x_3)|x_1-z|^2} \blp \eta^{jl}( \frac{k_3^{\mu}-k_2^{\nu}}{(x_3-z)(x_2-\bar{z})} -
\frac{k_2^{\mu}-k_3^{\nu}}{(x_2-z)(x_3-\bar{z})} )
+ (k_2 k_3)( \frac{\eta^{l \mu} \eta^{j \nu}}{(x_3-z)(x_2-\bar{z})} - \frac{\eta^{l \nu} \eta^{j \mu}}{(x_2-z)(x_3-\bar{z})}) \brp]
\nonumber \\
&-(2 p^i p^j (\eta^{l \mu} k_3^{\nu}- \eta^{l \nu} k_3^{\mu}) - \frac{1}{3}p^i p^j p^l \eta^{\mu \nu})
\frac{(z-\bar{z})^2}{|x_1-z|^2 |x_2-z|^2 |x_3-z|^2}  + (cyclic \: permutation \: of \: i,j,l)  \}\nonumber \\
& |x_1-x_2|^{4k_1 k_2}|x_1-x_3|^{4k_1 k_3}|x_2-x_3|^{4k_2 k_3} (x_1-z)^{2k_1 p} (x_1-\bar{z})^{2k_1 D p} \nonumber \\
& (x_2-z)^{2k_2 p} (x_2-\bar{z})^{2k_2 D p}(x_3-z)^{2k_3 p} (x_3-\bar{z})^{2k_3 D p} (z- \bar{z})^{pDp} \nonumber \\
&\times C_{D_2}^{ghost}\frac{(x_1 - x_2)(x_1 - x_3)(x_2 - x_3)}{z-\bar{z}}
\end{eqnarray}
We choose to fix the open string operator coordinates at $ x_1=
\infty, \, x_2= 1 \, ,x_3=0 $. To simplify the final expression we
use the symmetry of the graviton polarization tensor. The
expression to be integrated over the closed string position on the
world-sheet is :
\begin{eqnarray} \label{AmpIntegr}
&A \displaystyle \sim \int_{\cal{H}^{+}} d^2 \!z  \{ -u Tr(\epsilon D) (\ze_1 p) (\ze_2 \ze_3) + Tr(\epsilon D) ( - (\ze_1 \ze_2)(\ze_3 p)
\nonumber \\
 &-4(s+1)(\ze_1 \ze_2)(\ze_3 \epsilon k_3) ) \frac{1}{|z|^2} + 4t \frac{(\ze_1 \ze_2)(\ze_3 \epsilon k_2)}{|1-z|^2} \nonumber \\
 &(-4t (\ze_1 \ze_2)(\ze_3 \epsilon k_2) + 4u(\ze_1 \ze_2)(\ze_3 \epsilon k_1)) \frac{ (\frac{z + \bar{z}}{2})}{|z|^2|1-z|^2} \\
 &+(\ze_1 p)(16 (\ze_2 \ze_3)(k_2 \epsilon k_3) + 4u (\ze_2 \epsilon \ze_3) - \frac{1}{12} (\ze_2
 p)(\ze_3p) \nonumber \\
 &-16 (\ze_2 p)(\ze_3 \epsilon k_3))\frac{(\frac{z - \bar{z}}{2})^2}{|z|^2 |1-z|^2} \nonumber \\
 &+ (cyclic \: permutation \: of \: (k_{I},\ze_{I})) \} \nonumber \\
 &|1-z|^{s+u} |z|^{t+u} (z-\bar{z})^{-s-t-u-1} + (k_2, \ze_2) \leftrightarrow (k_3, \ze_3)\nonumber
\end{eqnarray}
 We use formula (\ref{FinInteg}) to perform the integrals. Permuting
the scalars allows us to combine several terms together and
eliminate undesired poles for $ s,t,u = -1 $ which appear in the
amplitude. So the string amplitude takes the final form :
\begin{eqnarray} \label{FinalRes}
&A = \displaystyle {\cal N} \frac{\Gamma(1-s)\Gamma(1-t)\Gamma(1-u)\Gamma(-
\frac{s+t+u}{2})}{\Gamma(1- \frac{t+u}{2})\Gamma(1-
\frac{s+u}{2})\Gamma(1- \frac{t+s}{2})\Gamma(1-
\frac{s}{2})\Gamma(1- \frac{t}{2})\Gamma(1- \frac{u}{2})}
\nonumber \\
&\times F(1,2,3,\epsilon)
\end{eqnarray}
where ${\cal N }$ is a normalization constant to be determined and
$F(1,2,3,\epsilon)$  is a form factor depending on the
polarizations and momenta:
\begin{eqnarray} \label{FormFac}
&F(1,2,3,\epsilon) = (s+u)(s+t) (\al_{1}+ \frac{\al_2}{2}) \nonumber \\
&+t(s+t)\al_3 +u(s+u)\al_4 +(s+t+u) ( s\al_5 + \al_6)  \nonumber \\
&+ (cyclic \: permutation \: of \: (k_{I},\ze_{I}))
\end{eqnarray}
The coefficients $ \alpha_{i} $ are given by:
  \begin{eqnarray}
  &\alpha_1= \frac{1}{2} (\ze_1 \ze_2)(\ze_3 p) Tr(\epsilon D), \qquad \alpha_2= 2 (\ze_1\ze_2)(\ze_3\epsilon k_3), \cr
  &\alpha_3= -2 (\ze_1 \ze_2)(\ze_3 \epsilon k_2), \qquad \alpha_4= -2 (\ze_1 \ze_2)(\ze_3 \epsilon k_1) , \cr
  &\alpha_5= 2(\ze_3 p) (\ze_1 \epsilon \ze_2), \\
  &\alpha_6= 8 (\ze_3 p) (\ze_1 \ze_2) (k_1 \epsilon k_2) -8(\ze_1 p)
  (\ze_2 p)(\ze_3\epsilon k_3) -\frac{2}{3}(\ze_1 p)(\ze_2 p)(\ze_3
  p) Tr(\epsilon D), \nonumber
  \end{eqnarray}

This amplitude contains poles which can be understood by
factorizing the world sheet as in (Fig.\ref{factorization}). The
poles appear when open string vertex operators collide on the
boundary of the disc. When this happens, open strings can
propagate as intermediate states between the two world sheets in
(Fig.\ref{factorization}).

\EPSFIGURE[ht]{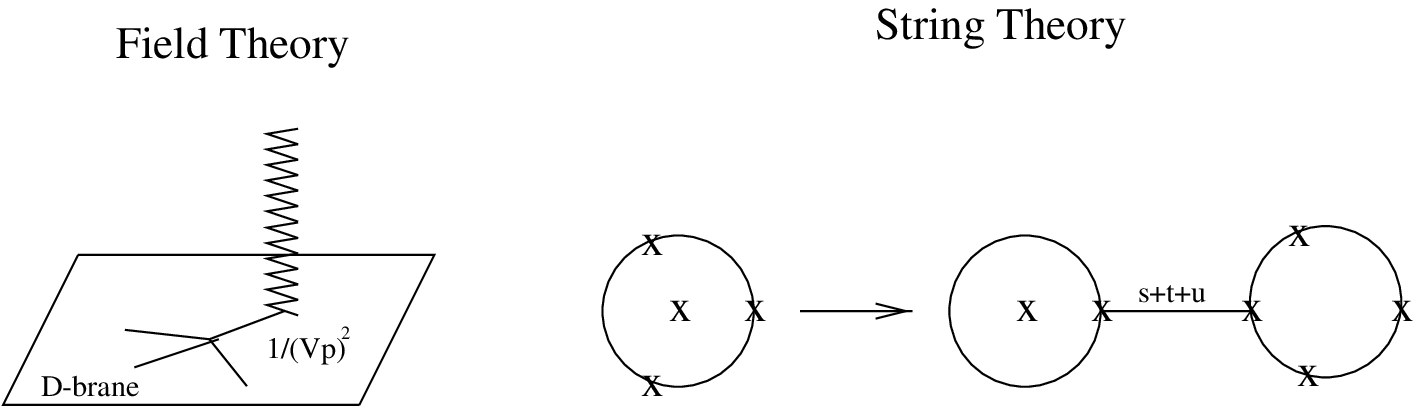}
{Factorization of the world-sheet giving rise to poles
from the exchange of open string modes between the two
world-sheets. The equivalent field theory diagram for small
momenta (massless pole for $(Vp)^2=0$). The wavy lines represent
gravitons and the plain ones open strings.\label{factorization}}

\section{Comparison with Born-Infeld Action}\label{normalization}
In order to determine the ($ \alpha^{'} $) corrections to the
Born-Infeld action (\ref{DBIWZ}) we will have to find the
normalization constant ${\cal N}$ of our amplitude. We will expand
the string amplitude for small momenta and compare the leading
contribution with the field theory scattering amplitude from the
Born-Infeld action. This will provide with a check of our
calculation as well. We follow the methods and conventions of
\cite{GarBI}. The world-volume theory of the D-brane includes the
massless fields $ X^{\mu} (\sigma) \ \ and \ \
A^{\alpha}(\sigma)$. The fields $ X^\mu(\sigma)$, where
$\sigma^{\alpha}$ the world-volume coordinates, are the embedding
of the p-brane in the ambient space-time. The fields
$A^{\alpha}(\sigma)$ are the gauge fields of the U(1) abelian
gauge theory on the brane. In this section we only consider the
abelian case which corresponds to a single brane dynamics. In
addition there are supersymmetric partners of those fields but
they are irrelevant to our case. The low energy dynamics of the
brane are encoded in the DBI-WZ action:
\begin{eqnarray} \label{DBIWZ}
&S_{D_p}= T_{(p)} \int dx^{p+1} e^{- \phi} \sqrt{ det \blp (G_{\mu
\nu}+ B_{\mu \nu}) \partial_{\alpha} X^{\mu} \partial_{\beta}
X^{\nu}
+ 2 \pi \apr F_{\alpha \beta} \brp} \nonumber \\
&-i T_{(p)} \int_{p+1} exp( 2\pi \apr F +B) \wedge \sum_{q} C_q
\end{eqnarray}
where $F_{\al \beta}$ the gauge boson field strength and the
$q$-form potentials $C_q$ with $q$ odd for type-$IIA$ and even for
type-$IIB$ are the Ramond-Ramond background fields.

 In the first step we will expand the Born-Infeld action to get the contact terms involving
three scalars and one graviton. We work in the Einstein frame with metric $ G_{\mu \nu}= e^{ \frac{\Phi}{2}}
g_{\mu \nu} $ . This is necessary because in this frame the bulk action for the
graviton takes the Hilbert-Einstein form $ (\sqrt{-g})^{\frac{1}{2}} \frac{{\bf R }}{2 \kappa^2} $ where
$ {\bf R} $ the bulk Ricci tensor and $\kappa$ the gravitational constant. In this frame the DBI action
takes the form
\begin{eqnarray} \label{EinsteinDBI}
S_{DBI}= T_{p} \int d^{p+1} \! \sigma \, Tr \blp e^{\frac{p-3}{4} \Phi }
\sqrt{ -det(\tilde{g}_{\alpha \beta} + e^{-\frac{\Phi}{2}}
\tilde{B}_{\alpha \beta} + 2 \pi l_{s}^2 e^{-\frac{\Phi}{2}}F_{\alpha \beta})}
\brp
\end{eqnarray}
where $ \tilde{g}_{\alpha \beta}$ , $ \tilde{B}_{\alpha \beta} $
are the pull-backs on the brane of the corresponding bulk tensors
\begin{eqnarray}\label{GPB}
\tilde{g}_{\alpha \beta}= g_{\mu \nu} \partial_{\alpha}X^{\mu} \partial_{\alpha}X^{\mu}
\end{eqnarray}
and similar expression for $ \tilde{B}_{\alpha \beta} $.

We work in the static gauge where the position of the p-brane is
fixed in the transverse dimensions. The world-volume coordinates
coincide with the bulk coordinates $ x^{\mu} $ for $ \mu = 0,
\dots p \ \ $(see Appendix \ref{submanifold}) which implies
\begin{eqnarray}
\partial_\alpha X^{\beta}= \delta_{\alpha}^{\beta} \nonumber \\
g_{\mu \nu}(X)=g_{\mu \nu}(\sigma,X^i)
\end{eqnarray}
with $\sigma$ the world-volume coordinates. The
fields $ X^i $ describe transverse fluctuations of the brane.
In this gauge (\ref{GPB}) takes the form
\begin{eqnarray} \label{GPB2}
\tilde{g}_{\alpha \beta}= g_{\alpha \beta} + 2 g_{i(\alpha}
\partial_{\beta)} X^i + g_{ij}\partial_{\alpha}X^i
\partial_{\beta}X^j
\end{eqnarray}

We want to expand for fluctuations around a flat empty space i.e.
$ g_{\mu \nu} = \eta_{\mu\nu}, B_{\mu \nu} = \Phi = 0 $, therefore
we have
\begin{eqnarray} \label{Fluc}
&g_{\mu \nu}= \eta_{\mu \nu} + 2 h_{\mu \nu}(\sigma,X^i)\nonumber \\
&B_{\mu \nu}= -2 b_{\mu \nu}(\sigma,X^i) \nonumber \\
&\Phi= \sqrt{2} \phi
\end{eqnarray}
where we have set $ \kappa =1$.
Applying (\ref{GPB2}) and (\ref{Fluc}) the Born-Infeld action
takes a simple form and we are able to expand the
square root of the determinant using the formula
\begin{eqnarray}
\sqrt{det(\delta_{\; \beta}^{\alpha}+ M_{\; \beta}^{\alpha})}= 1 + \frac{1}{2}
 M_{\; \alpha}^{\alpha}-\frac{1}{4}  M_{\; \beta}^{\alpha} M^{\beta}_{\; \alpha}
 +\frac{1}{8}  (M_{\; \alpha}^{\alpha})^2 + \dots
\end{eqnarray}

The three scalars and one graviton couplings from the DBI are
\begin{eqnarray}\label{IntLang}
&L_{contact}=  T_{p} \{ -( \partial _{\alpha} X^{i}\partial
^{\beta} X_{i} \partial ^{\alpha} X^{j} h_{j \beta} +
\partial ^{\alpha} X^{i}\partial ^{\beta} X_{i} \partial _{\beta}
X^{j} h_{j \alpha}) \nonumber \\
&+\partial_{\alpha}X^{i} \partial ^{\alpha}
X_{i} \partial^{\beta} X^{j} h_{j \beta}  \\
&\frac{1}{6} X^{i}X^{j}X^{k} \partial_{i} \partial_{j}
\partial_{k} h_{\alpha}^{\alpha} + X^{j} X^{k} \partial_{j}
\partial_{k} h_{ i \alpha} \partial^{\alpha}X^{i} \nonumber \\
&+ \frac{1}{2} X^{i} (\partial_{i} h_{\alpha}^{\alpha}) (\partial
X )^{2} - X^{j} (\partial_{j} h^{\alpha \beta})
\partial_{\alpha}X^{i} \partial_{\beta}X_{i} + X^{k}( \partial
_{k} h ^{i j} )
\partial_{\alpha}X_{i}\partial^{\alpha}X_{i} \}\nonumber
\end{eqnarray}
where the first three terms come from direct expansion of the
square root and the remaining from Taylor expansion of the
graviton in the transverse space, in contact terms with zero, one
and two scalars respectively. i.e
\begin{eqnarray}
&\frac{1}{2}M_{\; \alpha}^{\alpha}= h_{\;
\alpha}^{\alpha}(\sigma,X) + \dots =
h_{\; \alpha}^{\alpha}(\sigma,X)|_{X=0} + X^i \partial_ih_{\; \alpha}^{\alpha}(\sigma,X)|_{X=0} \\
&+ \frac{1}{2} X^i X^j \partial_i \partial_j h_{\;
\alpha}^{\alpha}(\sigma,X)|_{X=0}+\frac{1}{6} X^i X^j X^k
\partial_i
\partial_j \partial_k h_{\; \alpha}^{\alpha}(\sigma,X)|_{X=0} + \dots
\nonumber
\end{eqnarray}

Using for the scalar and graviton fluctuations the following
functional form
\begin{eqnarray} \label{FlucForm}
h_{\mu \nu}(\sigma,X^i)= \epsilon_{\mu \nu} e^{i (p^{\alpha}\sigma_{\alpha}+p^i X_i)}, \qquad
X^i(\sigma)= \ze^i e^{i k \cdot \sigma}
\end{eqnarray}
we can write down the Feynman rules for the vertices.

The polarizations $ \epsilon_{\mu \nu} \ \ and \ \ \ze^i $ satisfy
the same conditions as in the string amplitude (\ref{EqnMot}) in
section \ref{1cl3o}. The contribution of the interactions
(\ref{IntLang}) to the one graviton with three scalars  amplitude
is
\begin{eqnarray}\label{Acontact}
&A^{contact}= T_{p} \{ -\frac{(\ze_1 \ze_2)}{2}(s
(\ze_3 \epsilon k_3) -t (\ze_3 \epsilon k_2)-u(\ze_3
\epsilon k_1))  \\
&+\frac{1}{3}(\ze_1 p)(\ze_2 p)(\ze_3 p)Tr(\epsilon V) + 2(\ze_1
 p)(\ze_2 p)(\ze_3 \epsilon k_3) -\frac{s}{4}(\ze_3
p)(\ze_1
\ze_2) Tr(\epsilon V) \nonumber \\
&- 2(\ze_3 p)(\ze_1 \ze_2) (k_1 \epsilon k_2) -\frac{s}{2} (\ze_3
p)(\ze_1 \epsilon \ze_2) + (cyclic \: permutation \: of \:
(k_{I},\ze_{I})) \} \nonumber
\end{eqnarray}

These terms alone do not reproduce the string amplitude. As we can
see from the full expression in section \ref{1cl3o}, the amplitude
contains a pole (see Fig.\ref{factorization}) of the form $
\frac{1}{q^2} $, where $ q^2=-\frac{s+t+u}{2}$, see (\ref{Kinem}).
This pole is due to the exchange of a virtual scalar particle
between the graviton and the three open strings, see
(Fig.\ref{amplitude}). To construct these exchange terms we
need the graviton-scalar mixing and four scalar vertices from the
DBI.

\EPSFIGURE[ht]{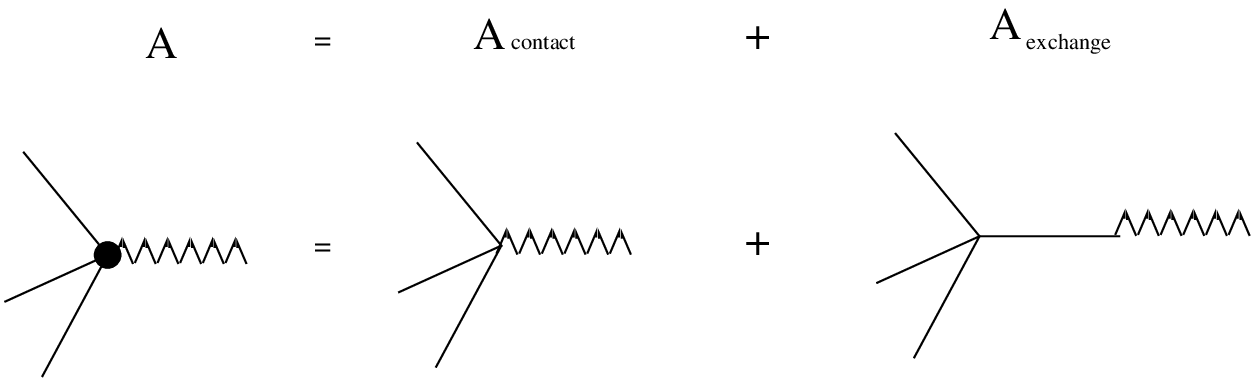}
{The field theory amplitude $A$ is the sum of a contact
part, $A^{contact}$ and a scalar exchange part, $A^{exchange}$.
\label{amplitude}}

\EPSFIGURE[ht]{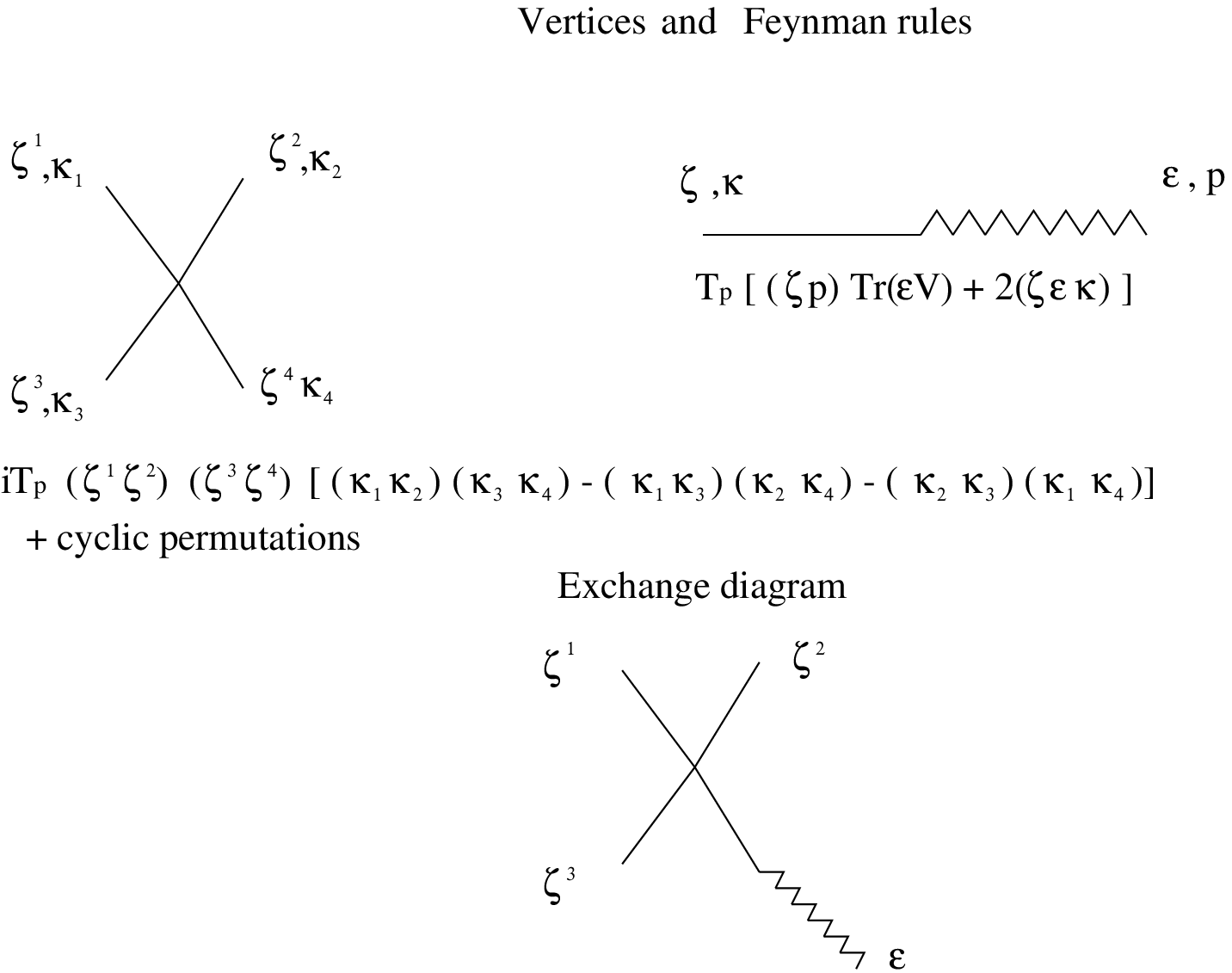}
{Feynman rules for the field theory calculation and the
exchange diagram for the 3scalar+1graviton scattering process.
\label{Feynman}}

\begin{eqnarray} \label{BIvert}
&L_{(X^4)}=T_{p}( -\frac{1}{4}\partial _{\alpha} X^{i}\partial
_{\beta} X_{i} \partial ^{\alpha} X^{j} \partial ^{\beta} X_{j} +
\frac{1}{8} \partial _{\alpha} X^{i}\partial ^{\alpha} X_{i}
\partial ^{\beta} X_{j} \partial _{\beta} X^{j}) \\
&L_{(Xg)}= T_{p}( X^{i} \partial _{i} h^{\alpha}_{\alpha} +
2\partial^{\alpha} X^{i} h_{\alpha i} ) \nonumber
\end{eqnarray}

These vertices are shown in (Fig.\ref{Feynman}) along with the
exchange diagram which contributes to our scattering amplitude.
 The field theory exchange amplitude can be written
\begin{eqnarray} \label{MixDiag}
A_{q^{2}}^{exchange} = V_{(X^4)} ^{i} P_{i}^{j} V_{j}^{(Xg)}
\end{eqnarray}
where $P_{i} ^{j} = \frac{ i \delta _{i}^{j}}{q^{2}}\frac{1}{T_p}$ 
is the propagator of a scalar field with momentum q and the two
vertices are the four scalar and scalar/graviton mixing vertices
,respectively, from the DBI terms above. Notice that unlike
\cite{GarBI} we have not absorbed $T_p$ in our normalization of
the scalars ( see (\ref{FlucForm})), which accounts for the
appearance of $T_p$ in the scalar propagator. The final expression
for $ A_{q^{2}}^{exchange}$ is
\begin{eqnarray}
A_{q^2}^{exchange}=\frac{T_p (\ze_1 \ze_2) t u}{8 q^2} \{ (\ze_3
p) Tr(\epsilon V) + 2 (\ze_3 \epsilon (k_1+k_2+k_3)) \}
\end{eqnarray}

The final amplitude is $ A_{FT}= A^{contact}+ A_{q^2}^{exchange}$
(see Fig.\ref{amplitude}). In order to compare the field theory
amplitude with the string amplitude, we have to take the $\alpha'
\to 0$ limit of the string amplitude. We restore $\alpha' $ in our
expression (\ref{FinalRes}) and we get
\begin{eqnarray} \label{aFinalRes}
&A= \displaystyle {\cal N} \frac{\Gamma(1-\frac{\alpha's}{2})\Gamma(1-\frac{\alpha' t}{2})\Gamma(1-\frac{\alpha' u}{2})\Gamma(-
\frac{\alpha'}{4}s+t+u)}{\Gamma(1- \frac{\alpha'}{2}t+u)\Gamma(1-
\frac{\alpha'}{4}s+u)\Gamma(1- \frac{\alpha'}{4}t+s)\Gamma(1-
\frac{\alpha's}{4})\Gamma(1- \frac{\alpha't}{4})\Gamma(1- \frac{\alpha'u}{4})} \nonumber \\
&\cdot F(1,2,3,\epsilon) = 
{\cal N }[ \frac{-4}{\alpha'(s+t+u)}+ {\cal O}(\alpha')] \cdot F(1,2,3, \epsilon)
\end{eqnarray}

Direct comparison of (\ref{aFinalRes}) to $A_{FT}$ verifies that
our result is correct and fixes the normalization constant
\begin{eqnarray}
{\cal N}= \frac{\alpha'T_{p}}{16}
\end{eqnarray}

\section{Derivative corrections to D-brane action I}
\label{corrections1}

The (${\cal O}(\alpha^{'2}) $) derivative terms involve squares of
pull-backs  to the normal and tangent bundle of the Riemann tensor
$R_{\mu \nu \rho \sigma}$. Their form is constrained by space-time
and world-volume reparametrization invariance. There are also
terms which involve the second fundamental form
$\Omega^{\mu}_{\alpha \beta}$ of the hyperplane. The explicit form
of these invariants as well as the definition of the second
fundamental form are given in Appendix \ref{submanifold}. In
\cite{BachasCurvature} the (${\cal O}(\alpha^{'2}) $) corrections
to the Born-Infeld action were determined by expanding the two
closed, one closed with two open and four open string amplitudes
to the subleading order in the momenta and comparing these with
the set of allowed derivative terms of the form $R^2, \, R
\Omega^2 $ and $ \Omega^4$ respectively. The comparison with the 2
closed strings amplitude determines $ R^{2} $ corrections to the
DBI modulo terms involving the bulk Ricci tensor $ R_{\mu \nu}$.
The reason terms proportional to $R_{\mu \nu} $ cannot be
determined, is that this tensor vanishes when we impose the lowest
order (in $ \alpha^{'}$) equations of motion for the graviton
(\ref{EQMotion}). There is only one ambiguity left for the $ R ^2
$ terms, which is proportional to the Gauss-Bonnet term
\begin{eqnarray}
L_{GB}=\frac{\sqrt{\tilde{g}}}{32 \pi^2} \blp R_{\alpha \beta \gamma
\delta }R^{\alpha \beta \gamma \delta } -4 \hat{R}_{\alpha \beta
}\hat{R}^{\alpha \beta }+ \hat{R}^2 \brp
\end{eqnarray}
where the Riemann tensors are the pull-backs to the world-volume
of the corresponding bulk tensors and $ \hat{R}_{\alpha \beta}, \
\ \hat{R} $ are obtained by contracting world-volume indices only
(see Appendix \ref{submanifold}). The Gauss-Bonnet term is a total
derivative at linear expansion of the Riemann tensors as can be
explicitly checked. This ambiguous coefficient was fixed in
\cite{BachasCurvature} by using IIA/F-theory duality. It was found
that it is vanishing. Nevertheless we will keep the coefficient of
this term as undetermined in our present analysis. As we shall
shortly explain there are additional ambiguities in the proposed
lagrangian. Through our amplitude calculations we will be able to
relate the coefficients of these ambiguous terms among themselves.
We will consequently use the conjectured duality to fix the values
of all these coefficients.

  There are also higher derivative terms involving the second
fundamental form $ \Omega $ of the hyperplane. Since the second
fundamental form is a scalar field excitation to linear order, we
can use the one closed with two open and four open string
amplitudes to determine the coefficients of $ R \Omega^2$ and $
\Omega ^4$ corrections to the lagrangian. In
\cite{BachasCurvature} it was found that the string amplitudes
mentioned above are reproduced by the lagrangian
  \begin{eqnarray} \label{LBachas}
&L^{(p)} =  T_{p} \, e^{-\phi} \sqrt{\tilde{g}} [ 1-\frac{1}{24} \frac{ (4
\pi^2 \alpha ^{'} ) ^{2} }{ 32 \pi ^{2}} ( (R_{T})_{\alpha \beta
\gamma
\delta }(R_{T})^{\alpha \beta \gamma \delta } \nonumber \\
&-2(R_T)_{\alpha \beta }(R_T)^{\alpha \beta }-(R_{N})_{\alpha
\beta i j }(R_{N})^{\alpha \beta i j } +2 \bar{R}_{i j} \bar{R}^{i
j})]
\end{eqnarray}
The tensors $ R_{T}$ and $ R_{N}$ are constructed from the
world-volume and normal bundle connections (see Appendix
\ref{submanifold}). In analogy with the $R^2$ terms, invariants
involving the trace of the second fundamental form, which to
linear order is the equation of motion for the scalar field
(\ref{EQMotion}), vanish and therefore cannot be determined  by
comparison with the above mentioned amplitudes.

Beyond these ambiguities we have again combinations of the $ R
\Omega^2$ and $ \Omega ^4$ which turn out to be total derivatives
for linear expansion in the fields of the tensors. In the
following formula we summarize the results found by
\cite{BachasCurvature} and the ambiguities involved. As explained
in Appendix \ref{submanifold} we have written their expression in
a slightly different form using (\ref{GaussCod}). We have also
included terms proportional to the trace of $ \Omega$. These terms
are non-vanishing when we expand the second fundamental form
beyond the leading order to get a graviton field.
\begin{eqnarray} \label{CurvBI}
&L^{(p)} =  T_{p} \, e^{-\phi} \sqrt{\tilde{g}} [ 1-\frac{1}{24} \frac{ (4
\pi^2 \alpha ^{'} ) ^{2} }{ 32 \pi ^{2}} ( (R_{T})_{\alpha \beta
\gamma
\delta }(R_{T})^{\alpha \beta \gamma \delta } \\
&-2R_{\alpha \beta }R^{\alpha \beta }-4R_{\alpha \beta }
(\Omega^{\alpha \gamma} \cdot \Omega_{\gamma}^{\;\beta})
-2(\Omega_{\alpha\gamma} \cdot  \Omega^{\alpha\delta})
(\Omega^{\beta\gamma} \cdot  \Omega_{\beta\delta}) \nonumber \\
&-(R_{N})_{\alpha \beta i j }(R_{N})^{\alpha \beta i j } +2
\bar{R}_{i j} \bar{R}^{i j} + L^{ambig})+ {\cal O}( \alpha^{'4}) ]
\nonumber
\end{eqnarray}
where $\bar{R}_{i j} \equiv \hat{R}_{ij} + g^{\alpha \alpha^{'}}
g^{\beta \beta^{'}} \Omega_{i|\alpha \beta} \Omega_{j|\alpha^{'}
\beta^{'}}$ defined in \cite{BachasCurvature} and we have included all the
ambiguities in the last term of the above expression. Using
(\ref{TotalDer},\ref{Ambiguities}) this lagrangian term takes the
form
\begin{eqnarray}\label{Lambi}
&L^{ambig}= \displaystyle  [ c_1\blp R_{\alpha \beta \gamma
\delta }R^{\alpha \beta \gamma \delta } -4 \hat{R}_{\alpha \beta
}\hat{R}^{\alpha \beta }+ \hat{R}^2 \brp \nonumber \\
&+c_2( 4 R_{\alpha \beta \gamma \delta} (\Omega^{\alpha \gamma}
\cdot \Omega ^{\beta \delta}) -8 \hat{R}_{\alpha \beta}
(\Omega^{\alpha \gamma} \cdot \Omega ^{\;\beta}_{\gamma})+2
\hat{R}(\Omega^{\alpha
\beta} \cdot \Omega _{\alpha \beta}) ) \nonumber \\
&+ c_3 (2(\Omega_{\alpha \gamma} \cdot \Omega_{\beta \delta})
(\Omega^{\alpha \gamma} \cdot \Omega^{\beta \delta}) -2
(\Omega_{\alpha \gamma} \cdot \Omega_{\beta \delta})
(\Omega^{\alpha \delta} \cdot \Omega^{\beta \gamma}) \nonumber \\
&-4(\Omega_{\alpha \gamma} \cdot \Omega_{\beta}^{\; \gamma})
(\Omega^{\alpha \gamma} \cdot \Omega^{\;\beta}_{\gamma}) +
(\Omega^{\alpha \beta} \cdot \Omega_{\alpha \beta})(\Omega^{\gamma
\delta} \cdot \Omega_{\gamma \delta}) )\nonumber \\
&+ c_4((\Omega_{\gamma}^{\; \gamma} \cdot
\Omega_{ \alpha}^{\; \beta})(\Omega^{\; \alpha}_{\delta} \cdot
\Omega_{\beta}^{\; \delta})) +c_5(\hat{R}_{\alpha \beta}
(\Omega^{\alpha \beta} \cdot \Omega ^{\; \gamma}_{\gamma}))]
\end{eqnarray}
These are all the ambiguities, involving one or no trace of the second fundamental form.

In \cite{BachasCurvature} it was guessed that the $ R \Omega ^2 $
terms in the second line of $L^{(p)}$ can be written in terms of
invariants constructed from $ (R_{T})_{\alpha \beta}$ as in
(\ref{LBachas}). Using the Gauss-Codazzi equations
(\ref{GaussCod}) and the definition $(R_T)_{\alpha \beta}=
(R_T)^{\gamma}_{\; \alpha \gamma \beta}$ these terms become
\begin{eqnarray}
&(R_T)_{\alpha \beta }(R_T)^{\alpha \beta }= R_{\alpha \beta
}R^{\alpha \beta }+2R_{\alpha \beta } (\Omega^{\alpha \gamma}
\cdot \Omega_{\gamma}^{\;\beta}) +(\Omega_{\alpha\gamma} \cdot
\Omega^{\alpha\delta})
(\Omega^{\beta\gamma} \cdot  \Omega_{\beta\delta}) \nonumber \\
&-2(\Omega_{\gamma}^{\; \gamma} \cdot \Omega_{ \alpha}^{\;
\beta})(\Omega^{\; \alpha}_{\delta} \cdot \Omega_{\beta}^{\;
\delta})-2\hat{R}_{\alpha \beta} (\Omega^{\alpha \beta} \cdot
\Omega ^{\; \gamma}_{\gamma})+ (\Omega_{\gamma}^{\; \gamma} \cdot
\Omega_{ \alpha}^{\; \beta})(\Omega_{\gamma}^{\; \gamma} \cdot
\Omega_{ \alpha}^{\; \beta}) \nonumber
\end{eqnarray}
 It is obvious from the equation above that if this were the case
 then the terms proportional to $ c_4 $ and $ c_5$ should appear.
We will see that these terms cannot be determined through our
string amplitude methods and therefore remain ambiguous in the
effective lagrangian. In addition comparison with our string amplitudes
will show that the ambiguous terms combine to form the
Gauss-Bonnet lagrangian for the induced metric $g_{\al \beta}$
up to terms proportional to the trace of the second fundamental form.

Now we proceed with the Field Theory computation. Contact terms
involving one closed and three open strings come from pull-backs
and Taylor expansion in the graviton of the $ R \Omega ^{2} $
terms, as well as from $ \Omega ^4$ terms from expansion of the
second fundamental form beyond the leading order. It is obvious
that the coefficients $ c_1$ and $ c_5$ cannot be determined by
employing the three scalars and one graviton string amplitude.
Both terms involve at least two gravitons and we will need the two
graviton with one scalar amplitude to fix them. As an example of
the expansion needed for the computation we  write down the
pull-back of one $R \Omega ^{2} $ term keeping only the pieces
relevant to our case
\begin{eqnarray} \label{PullBack}
R^{\alpha \beta }_{\ \ \ i j} \Omega^{i}_{\alpha \gamma} \Omega^{j
\gamma}_{ \beta} \to
\partial_{\alpha} X_{\mu} R^{\mu \beta}_{\; \; i j} \Omega^{i}_{\alpha \gamma} \Omega^{j
\gamma}_{\beta} + \xi _{i}^{\mu} R^{\alpha \beta}_{\ \ \ \mu j}
\Omega^{i}_{\alpha \gamma} \Omega^{j \gamma}_{\beta} + (\alpha
\leftrightarrow \beta, i \leftrightarrow j)
\end{eqnarray}
In addition the second fundamental form in static gauge has the following expansion
\begin{eqnarray} \label{SecFundExp}
\Omega_{\alpha \beta} ^{i}= \partial_{\alpha} \partial_{\beta}
X^{i} + \Gamma_{\alpha \beta}^{i} + \dots 2-particle \: terms
\end{eqnarray}
The expansion of the Christoffel symbol around a flat background gives to lowest order
one graviton field.
\begin{eqnarray} \label{Christof}
\Gamma_{\alpha \beta}^{i} = [\partial_{\beta} h_{\alpha i}+\partial_{\alpha} h_{\beta i}
-\partial_{i} h_{\alpha \beta}]
\end{eqnarray}
 This expansion can be used for one of the second fundamental forms in
$ \Omega^4$ to give contact interactions of three scalars and one
graviton.

 A tedious calculation gives the contribution of the
contact terms to the field theory amplitude
\begin{eqnarray}
&A_{FT}^{contact}= \displaystyle -\frac{T_p (4 \pi^2 \alpha^{'})^2}{24 (32 \pi ^2)} \{(s^3 + u^2 s +t^2 s) \frac{\beta_1}{4}
+(s^2+t^2+u^2)(-\frac{\beta_2}{6} - \beta_3 + \frac{\beta_4}{8}) \nonumber \\
&+(s(s^2+t^2+u^2)-ut(s-u-t))\frac{\beta_5}{8} -(s^2u+stu+u^3-u^2t)\frac{\beta_6}{4} \nonumber \\
&- (s^2t+stu+t^3-t^2u) \frac{\beta_7}{4} \nonumber \\
&-(s^2(t+u)+4stu-2s^3) \frac{\beta_8}{8} + (c_2-c_3)[ u^2 \frac{\beta _9}{2} +
t^2 \frac{\beta _{10}}{2} \nonumber \\
&-ut \beta_4 -us \beta_{11} -ts \beta_{12} + s^2 \frac{\beta_{13}}{2}] \\
&sut[(-\frac{c_2}{4} + \frac{c_4}{32}) \beta_5 + ( -\frac{c_3}{2} + \frac{c_4}{16})
(\beta_6+\beta_7 + \beta_8)]
\nonumber \\
&+ (cyclic \: permutation \: of \: (k_{I},\ze_{I})) \} \nonumber
\end{eqnarray}
The coefficients $ \beta_i $ are functions of the polarizations and momenta
\begin{eqnarray}
&\beta_1= (\ze_1 \epsilon \ze_2) ( \ze_3 p), \qquad \beta_2=(\ze_1 p)(\ze_2 p)(\ze_3 p) Tr(\epsilon V),
 \nonumber \\
&\beta_3= (\ze_1 p)(\ze_2 p)(\ze_3 \epsilon k_3), \qquad \beta_4= (\ze_1 \ze_2)(\ze_3 p) (k_1 \epsilon k_2),
\nonumber \\
&\beta_5=(\ze_1 \ze_2)(\ze_3 p) Tr(\epsilon V), \qquad \beta_6=(\ze_1 \ze_2)(\ze_3 \epsilon k_3), \\
&\beta_7=(\ze_1 \ze_2)(\ze_3 \epsilon k_1), \qquad \beta_8=(\ze_1 \ze_2)(\ze_3 \epsilon k_2), \nonumber \\
&\beta_9=(\ze_1 \ze_2)(\ze_3 p )(k_1 \epsilon k_1), \qquad \beta_{10}=(\ze_1 \ze_2)(\ze_3 p )(k_2 \epsilon k_2),
 \nonumber \\
&\beta_{11}=(\ze_1 \ze_2)(\ze_3 p )(k_1 \epsilon k_3), \qquad \beta_{12}=(\ze_1 \ze_2)(\ze_3 p )(k_2 \epsilon k_3),
\nonumber \\
&\beta_{13}=(\ze_1 \ze_2)(\ze_3 p )(k_3 \epsilon k_3), \nonumber
\end{eqnarray}
$A_{FT}^{contact}$ does not reproduce the full string amplitude.
The previously computed amplitudes
\cite{GarScatter,BachasCurvature} had no poles to the subleading
order in the momenta because the trilinear bulk and
scalar-graviton mixing vertices are presumably protected by
supersymmetry and do not receive derivative corrections.
Nevertheless, as already mentioned above, the four scalars vertex
has derivative corrections, so one needs to include scalar
exchange diagrams of these vertices with the scalar-graviton
mixing vertex. This is in exact analogy with the DBI case of
section \ref{normalization}. Each $ \Omega ^4$  and $ \Omega^3 Tr \Omega$
term (with $Tr \Omega$ giving the off-shell exchanged scalar)
contributes a four scalars vertex $( \sim \, k^{8} )$ where k the
momentum of the scalars. The mixing diagrams formed between these
vertices and $ L_{(Xg)} $ from (\ref{BIvert}) are of the same
order as the contact terms. The explicit form of the exchange
diagram contribution is given by
\begin{eqnarray}
&A^{exchange}_{FT}=  \displaystyle -\frac{T_p (4 \pi^2 \alpha^{'})^2}{24 (32 \pi ^2)} \{ [\frac{ (s^2-ut)ut}{4(s+u+t)}
+ sut (\frac{c_3}{4} - \frac{c_4}{32})]  \\
&(\beta_5 +2 \beta_6 + 2 \beta_7 + 2\beta_8) + (cyclic \: permutation \: of \: (k_{I},\ze_{I})) \}
\nonumber
\end{eqnarray}

We now expand the three scalars and one graviton amplitude from
(\ref{aFinalRes}) to subleading order in momenta
\begin{eqnarray} \label{AmpExp}
A= -\frac{T_{p}}{8} [ \frac{2}{s+t+u}+ \frac{1}{24} \frac{(4 \pi^2
\alpha^{'})^2} {32 \pi^2} \frac{s^2 + t^2 +u^2}{s+t+u} +
{\cal O}(\alpha^{'4})] \cdot F(1,2,3,\epsilon)
\end{eqnarray}
The subleading piece of the expansion, as expected, has a pole
which is not canceled. The full field theory amplitude $
A_{FT}=A_{FT}^{contact} + A_{FT}^{exchange}$ (Fig.\ref{amplitude})
is compared to the string amplitude expansion. The two expressions
come to complete agreement for the following values of the unknown
coefficients
\begin{eqnarray}\label{ambiguous1}
c_2=c_3 \nonumber
\end{eqnarray}
while the exchange and contact contributions, from the lagrangian term
1 to $c_4$, cancel against each other. Therefore
\begin{eqnarray}
c_4 \to undetermined \nonumber
\end{eqnarray}

As we already saw the coefficients  $c_2,c_3$ which correspond to
terms that do not involve the trace of $ \Omega$ cannot be
completely fixed by our computations but they are proportional to
each other. On the other hand the coefficient $c_4$ of the
interaction term proportional to the trace of the second
fundamental form is completely undetermined. We will explain the
source of this remaining ambiguity in our conclusions.

In the next section we will try to determine the remaining
coefficients $c_1, c_5 $. The terms proportional to these
coefficients are non-vanishing for a two graviton with one scalar
amplitude. Therefore we will need to compute the string amplitude
of two closed strings with one open string excitation.

\section{Scalar-two graviton scattering amplitude}
\label{2cl1o}

In this section we will determine the remaining ambiguities in $
L^{(p)}$. The terms we are interested in are proportional to $c_1
, \ c_2 \ and \ c_5$. The $c_1$ and $c_5$ terms have an expansion
which contributes to amplitudes at least quadratic in the graviton
field. The two graviton scattering amplitude is not enough to fix
the $c_1$ term as explained in section \ref{corrections1}. The
next simplest case where such terms contribute involves two
gravitons and one scalar excitation. In addition the amplitude
receives contributions from the $c_2$ terms as well. In conclusion
we need to compute the string amplitude with two closed and one
open strings. Comparison of the low energy expansion of this
amplitude with $L^{(p)}$ will eventually fix some of the ambiguous
terms.

We could go on and compute the string amplitude using the
correlators given in section \ref{preliminaries} but it will prove
difficult to perform the integration over the position of string
vertex operators on the complex plane. On the other hand, we can
use correlators on the disc rather than the half complex plane.
The Green's functions on the disc are found using the method of
image charges on a two dimensional surface \cite{Burgess}. Each
string inserted at position $z$ on the disc has an image at
$\frac{1}{\bz}$. Imposing Neumann or Dirichlet boundary conditions
we find the correlators on the disc \cite{klebanov}:
\begin{eqnarray} \label{corrD}
&\langle X^{\mu}(z) X^{\nu}(w) \rangle= -\eta^{\mu \nu} ln (z-w)
\nonumber \\
&\langle X^{\mu}(z) \bar{X}^{\nu}(\bw) \rangle= -D^{\mu \nu} ln
(1-z \bw) \nonumber \\
&\langle \psi^{\mu}(z) \psi^{\nu}(w) \rangle = \displaystyle - \frac{\eta^{\mu
\nu}}{z-w} \nonumber \\
& \langle \psi^{\mu}(z) \bar{\psi}^{\nu}(\bw) \rangle = \displaystyle + i
\frac{D^{\mu \nu}}{1-z \bw}
\end{eqnarray}
From the form of the correlators we see that we can still
use the substitutions (\ref{substitute}) for right movers in terms
of left movers with anti-holomorphic arguments. Holomorphic and
anti-holomorphic operators have non-trivial correlators given by
(\ref{corrD}). For the superghost fields we find:
\begin{eqnarray} \label{sgcorrD}
\langle e^{-\phi(z)} e^{-\phi(\bw)} \rangle= \frac{1}{1-z \bw}
\end{eqnarray}
and the ghost correlators can be determined from those of the
sphere using the doubling trick and transformation properties of
the fields under the involution $z'= \bz^{-1}$ \cite{POLCHINSKI}:
\begin{eqnarray} \label{ghcorrD}
&\langle c(z_1) c(z_2) c(z_3) \rangle = C_{D_2}^{ghost}
(z_1-z_2)(z_2-z_3)(z_1-z_3) \nonumber \\
&\langle c(z_1) c(z_2) \bar{c}(\bz_3) \rangle= (\frac{\partial
z'_3}{\partial \bz_3})^{-1} \langle c(z_1) c(z_2) c(z'_3) \rangle
\nonumber \\
&=C_{D_2}^{ghost} (z_1-z_2)(1-z_1 \bz_3)(1-z_2 \bz_3)
\end{eqnarray}
Before we proceed with the string computation we will give some
useful formulae for computing the vertex operator correlators:
\begin{eqnarray} \label{corrformulae}
&\langle \partial_z X^{\mu}(z) e^{i (Dp) \cdot X(\bw)} \rangle =
\displaystyle \frac{i (Dp)^{\mu} \bw }{1-z \bw} \nonumber \\
&\langle \partial_{\bz} X^{\mu}(\bz) e^{i p \cdot X(w)} \rangle =
\displaystyle \frac{i (Dp)^{\mu} w }{1-\bz w} \nonumber \\
&\langle \partial_z X^{\mu}(z) \partial_{\bw} X^{\nu}(\bw)
\displaystyle \rangle= \frac{ \eta^{\mu \nu}}{(1-z \bw)^2}
\end{eqnarray}

The two closed with one open string amplitude has the general form
(\ref{StrAmp}). We choose to fix the positions of one closed and
one open string. We also put the fixed closed string vertex
operator in the $(-1,-1)$ picture. The string amplitude takes the
form:
\begin{eqnarray} \label{2cl+1o}
&A_{string} \simeq \displaystyle \int_{|z| \leq 1} d^2 \!z \, \langle c(z')
\bar{c}(\bz') V^{\mu \nu }_{(-1,-1)}(z, \bar{z}) \, V^{\rho
\si}_{(0,0)}(z,\bar{z}) c(x)V^i_{(0)}(x) \rangle \nonumber \\
&\times (\epsilon D)_{\mu \nu} (\beps D)_{\rho \si} \ze_i
\end{eqnarray}
where the vertex operators for the closed strings are given by the
expressions (\ref{PicOper}) and for open by equation
(\ref{OpenOper}). The momenta for the closed strings with
polarizations $\epsilon^{\mu \nu}$ and $\beps^{\rho \si}$ are
$p^{\mu}$ and $\bp^{\mu}$ respectively. The momentum of the open
string is $k^{\mu}$.

We introduce kinematic invariants:
\begin{eqnarray}\label{kinematics2}
&s=\bp \cdot D \cdot p \qquad t= \bp \cdot p \qquad u= \bp \cdot D \bp \\
&p \cdot D \cdot p= -2(s+t) -u \qquad 2p \cdot k = -2 \bp \cdot k= s+t+u \nonumber
\end{eqnarray}
In the last two equations we used the conservation of momentum
along the world-volume directions:
\begin{eqnarray}\label{conservation2}
(V \bp)^{\mu} + (Vp)^{\mu} +k^{\mu}=0
\end{eqnarray}
The gauge transversality conditions are the same as for the three
open with one closed string amplitude (\ref{EqnMot}).
The string amplitude in this case though, is far more more tedious
than before. We can simplify the calculations involved by
restricting the momenta and/or polarizations of one closed string
to lie on the D-brane directions. If we impose
this restriction for both momenta and polarizations of the closed
string ($\beps^{\mu \nu}, \bp^{\mu}$), then as it turns out, the amplitude does not encode
enough information to fix the ambiquous terms. 
On the other hand restricting only the polarization $\beps^{\mu \nu}$
on the world-volume is sufficient for our purpose. The gauge
transversality conditions for $\beps^{\mu \nu}$ become:
\begin{eqnarray}\label{restriction}
&\bp^{\mu} \beps_{\mu \nu}= \bp^{\al} \beps_{\al \beta}=0
\nonumber \\
&\beps^{\mu}_{\mu}=\beps^{\al}_{\al}=0
\end{eqnarray}

In addition, the ambiguous terms proportional to $c_1$ involve
pull-backs of the Riemann tensor with all the indices on the
world-volume. Contact interaction terms with two gravitons and one
scalar come from the pull-back of the Riemann tensor or the
Taylor expansion of the graviton in the Riemann tensor:
\begin{eqnarray}
&R_{\al \beta \gamma \de} \to X^{i} \partial_i R_{\al \beta \gamma
\de}+
\partial_{\al}X^{i} R_{i \beta \gamma \de}+ \dots \nonumber
\end{eqnarray}
where the terms omitted involve more than one scalars.
 In either case we have only one index of
the bulk Riemann tensor on the normal directions. A similar
discussion applies for the terms proportional to $c_5$. We can
therefore focus our attention on terms involving only one normal
index in the string amplitude. We need only to be careful in applying
the gauge transversality conditions (\ref{EqnMot}) to reduce any
normal indices to tangent where this is possible. With these two
restrictions the computation simplifies considerably and as we
shall see shortly there is enough information encoded in the
string amplitude to determine the ambiguous terms.
The integrand of the string amplitude (\ref{2cl+1o}) takes the
form:
\begin{eqnarray}\label{integrand2cl+1o}
&\langle \; : (\partial X^{\al} + i (\bp \cdot \psi)\psi^{\al})(z)
e^{i \bp \cdot X(z)}: \; :(\partial X^{\beta} + i ((D \bp) \cdot
\psi)\psi^{\beta})(\bz) e^{i(D \bp) \cdot X(z)}: \nonumber \\
&:c(x)(\partial X^i + 2i (k \cdot \psi) \psi^i)(x) e^{2i k \cdot
X(x)}: \\
&:c(z') e^{- \phi(z')} \psi^{\mu}(z') e^{ip \cdot X(z')}: :c(\bz')
e^{- \phi(\bz')} \psi^{\mu}(\bz') e^{i(Dp) \cdot X(\bz')} : \;
\rangle \; \epsilon_{\al \beta} (\epsilon D)_{\mu \nu} \ze_i
\nonumber
\end{eqnarray}
The integrand breaks into four different types of path integrals
to be evaluated in the same manner as in section \ref{1cl3o}. We
will fix the position of the $\epsilon_{\mu \nu}$ closed string at
the center of the disc $z'=\bz'=0$ and of the open string at
$x=1$.

The final result after the evaluation of all path integrals is:
\begin{eqnarray}\label{integrandfinal}
&\displaystyle \{ \frac{|1+z|^2}{|1-z|^2 |z|^2} \blp -Tr(\epsilon D) (p \beps
p)(\ze p) + 4 (\ze \epsilon p) (p \beps p) +2(\ze \epsilon \beps
p)(s+t+u)
\brp \nonumber \\
&\displaystyle \frac{1-|z|^2}{|1-z|^2 |z|^2} \blp -Tr(\epsilon D) (p \beps
p)(\ze \bp) + 4(\ze \bp) (p \beps \epsilon p) - (\ze \bp)
Tr(\epsilon \beps) (2s+2t+u) \brp \nonumber \\
& \displaystyle \frac{1}{|z|^2} ( -u Tr(\epsilon \beps) (\ze p)) +
\frac{z+\bz}{|1-z|^2 |z|^2}(-2u(\ze \epsilon \beps p)) \}
\nonumber \\
&|1-z|^{2(-u-t-s)} (1-|z|^2)^u |z|^{2t}
\end{eqnarray}
The integration over the closed string coordinate $z$ can be
performed using the formulas from Appendix \ref{integrals2}.
The final result is:
\begin{eqnarray}\label{stringresult2}
A_{string}= {\cal N} \{(2I_1+I_3) (-2 \gamma_1+4\gamma_2 +
2(s+t+u)\gamma_3 )+ \nonumber \\
I_2 (-2\gamma_4 + 4 \gamma_5 -(2s+2t+u) \gamma_6) + I_3(-u
\gamma_7) + I_1 (-2u \gamma_3) \}
\end{eqnarray}
where the expressions $I_1,I_2,I_3$ are defined in Appendix
\ref{integrals2} and ${\cal N}$ a normalization constant. We
have also defined:
\begin{eqnarray}\label{formfactor2}
&\gamma_1= Tr(\epsilon V) (p \beps p)(\ze p) \qquad \gamma_2=(\ze
\epsilon Vp)(p \beps p) \nonumber \\
&\gamma_3=(\ze \epsilon \beps p) \qquad \gamma_4= Tr(\epsilon V)
(p \beps p)(\ze \bp) \nonumber \\
&\gamma_5= (\ze \bp) (p \beps \epsilon p)
\qquad \gamma_6=(\ze \bp) Tr(\epsilon \beps)  \\
&\gamma_7= Tr(\epsilon \beps) (\ze p) \nonumber
\end{eqnarray}

The final result involves generalized hypergeometric functions.
Manipulating these functions to bring our result in a more compact
form is a difficult task. In addition for comparison with the low
energy effective action $L^{(p)}$ we should expand the
hypergeometric functions for small $s,t,u$. There is not such
known expansion of these functions. Fortunately we shall need
neither the normalization constant ${\cal N}$ nor the small
momenta expansion of (\ref{stringresult2}). The relative
coefficients of the various polarization contractions in the
expressions (\ref{formfactor2}) will be sufficient for determining
the unknown coefficients $c_1,c_5$ as well as for verifying that
our computation is indeed correct.

\section{Derivative corrections to D-brane action
II}\label{corrections2}

Now we proceed with the field theory computation. Contact terms
involving two gravitons and one scalar come from pull-backs and
Taylor expansion of $R^2$ terms as well as from $R \Omega^2$ terms
from expansion of the second fundamental form beyond the leading
order. Extracting these contact terms is a very tedious process.
As an example we give the contribution from one $R^2$ term:
\begin{eqnarray}\label{exampleR2}
&R_{\al \beta \gamma \de} R^{\al \beta \gamma \de} \to
\partial_{\al} X^{i} R_{i \beta \gamma \de} R^{\al
\beta \gamma \de} + \partial_{\al} X^{i} R_{i \beta \gamma \de}
R^{\al \beta \gamma \de}  \nonumber \\
&+ \ all \ other \ pull-backs \ + X^i \partial_i( R_{\al \beta
\gamma \de} R^{\al \beta \gamma \de}) \nonumber
\end{eqnarray}
The contribution of this term is:
\begin{eqnarray}
&R_{\al \beta \gamma \de} R^{\al \beta \gamma \de} \to \nonumber
\\
&2[ (\ze p) \blp Tr(\epsilon \beps)(-\frac{(s+t)(s+t+2u)}{4}) + (p
\beps \epsilon \bp) (s+t+u) - (p \beps p)(\bp \epsilon \bp) \brp
\nonumber \\
&(\ze \bp) \blp Tr(\epsilon \beps)(-\frac{(s+t)(3s+3t+2u)}{4}) -(p
\beps \epsilon \bp) (s+t+u) - (p \beps p)(\bp \epsilon \bp)
\nonumber \\
&+(p \beps \epsilon p) (s+t) -2 (p \beps p)(\bp \epsilon p) \brp
\nonumber \\
&+(\ze \epsilon \beps p) (\frac{(s+t)u}{2})-(\ze \epsilon \bp)(p
\beps p) u] \nonumber
\end{eqnarray}
where we have used the equations of motion and gauge
transversality conditions to reduce the result in terms of
appropriate invariants for comparison with the string amplitude.
For the $R \Omega^2$ terms we use the expansion of the second
fundamental form in static gauge (\ref{SecFundExp}) to get contact
interactions of two gravitons with one scalar.

The polarization tensor $\beps^{\al \beta}$ has only world-volume
indices and this simplifies the procedure considerably. As an
example there is no contribution from the $\hat{R}^2$ term since
at least one Ricci scalar will be expanded to linear order of the
graviton field $\beps^{\al \beta}$ and therefore it vanishes due
to the conditions (\ref{restriction}). As another example,
invariants like $(R_{\al \beta ij})^2$ contribute:
\begin{eqnarray}
&R_{\al \beta ij} R^{\al \beta ij} \to \partial_{\al} X^k R_{k
\beta ij} R^{\al \beta ij} + \xi^{\gamma}_i R_{\al \beta \gamma j}
R^{\al \beta ij} \nonumber \\
&+ \ all \ other \ pull-backs \  + X^i \partial_i(R_{\al \beta ij}
R^{\al \beta ij}) \nonumber
\end{eqnarray}
The normal frame and Taylor expansions vanish and only the tangent
frame pull-back contributes.

Again the contact terms alone do not reproduce the full string
amplitude. One has to consider exchange terms as well. The two
scalars with one graviton vertex of the Born-Infeld action has
derivative corrections $R \Omega^2$, so we need to include scalar
exchange diagrams as in (Fig.\ref{amplitude2}). Each $R \Omega^2$
term gives a vertex with two scalars and one graviton of sixth
order in the momenta. Combined with the mixing vertex from
$L_{(Xg)}$ in (\ref{BIvert}) we get exchange terms of the same
order as the contact terms. The general form of the exchange terms
is:
\begin{eqnarray}\label{exchange2}
A_{FT}^{exchange}= V_{Xg}^i V^j_{XXg} P_{ij}
\end{eqnarray}
where the vertex $V^j_{XXg}$ comes from expansion to linear order
of the Riemann and second fundamental form tensors in $R
\Omega^2$ and $R \Omega Tr \Omega$.
\EPSFIGURE[ht]{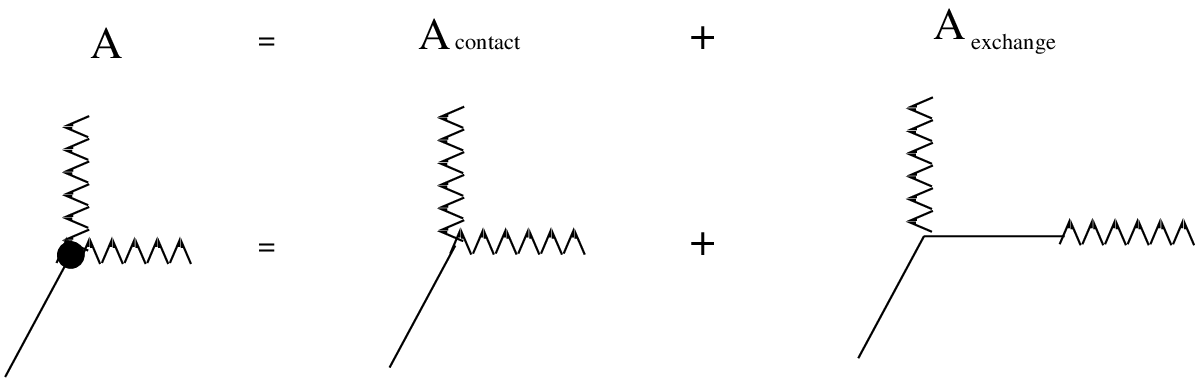}
{The ${\cal O}(\apr^2)$ field theory amplitude $A$ from $L^{(p)}$ 
is the sum of a contact part, $A^{contact}$ and a scalar exchange part,
$A^{exchange}$.\label{amplitude2}}

Combining all of the above contributions for the proposed
lagrangian $L^{(p)}$ we get:
\begin{eqnarray}\label{fieldtheory2}
&A_{FT} \sim \gamma_3(\frac{s^2 -t^2 -u^2}{2}) + -2 \gamma_2
(\frac{u^2}{2s+2t+u} + (t-s)) \nonumber \\
&+ \gamma_7 ( \frac{u^2}{4} + (c_2-c_1)\frac{(s+t+u)^2}{2}) +
\gamma_8(c_1-c_2)(2s+2t) + 2\gamma_{9}(c_1-c_2) \nonumber \\
&+ \gamma_1( \frac{u^2}{2s+2t+u} + (t-s)+ (c_2-c_1)u) + \gamma_{10} (c_2-c_1)(2u) \\
&\gamma_6( \frac{4(s+t)(s+t+u) + u^2}{4} + (c_1-c_2)
\frac{-2(s+t)^2-4u(s+t) -2u^2}{4} )  + 2\gamma_{11} (c_2-c_1)
\nonumber \\
&+\gamma_{12} (c_1-c_2)(2s+2t)+  \gamma_5(2s+2t+u +(c_2-c_1)2u) +
\gamma_4(-\frac{2s+2t+u}{2} + (c_1-c_2)u) \nonumber
\end{eqnarray}
where the form factors are given by (\ref{formfactor2}) and the
definitions:
\begin{eqnarray}\label{extraformfactor}
&\gamma_8= (\ze p) (p \beps \epsilon \bp) \qquad \gamma_{9}= (\ze
p) (p \beps p)(\bp \epsilon \bp) \nonumber \\
&\gamma_{10}=(\ze p)(p \beps \epsilon p) \qquad \gamma_{11}=(\ze
\bp) (p \beps p)(\bp \epsilon \bp) \\
&\gamma_{12}= (\ze \bp) (p \beps \epsilon \bp)
\end{eqnarray}

We observe that $c_5$ dropped out of the expression
(\ref{fieldtheory2}) since the contact piece canceled against the
exchange as in the case of $c_4$. We can determine $c_1$ and $c_2$
by comparison with the string amplitude (\ref{stringresult2}). The
extra form factors defined in (\ref{extraformfactor}) are absent
in the string result. The conditions for absence of these terms
are supplemented by the conditions imposed by the relative
coefficients of the form factors in (\ref{stringresult2}). As a
result, the field theory amplitude reproduces correctly the string
amplitude structure if the ambiguous coefficients satisfy the
relations:
\begin{eqnarray}\label{ambiguous2}
c_1=c_2 \nonumber
\end{eqnarray}
while as mentioned before:
\begin{eqnarray}
c_5 \to undetermined\nonumber
\end{eqnarray}

Combining the conditions (\ref{ambiguous1}) and (\ref{ambiguous2})
we have determined the ambiguous coefficients $c_2,c_3$ in terms
of only one unknown $c_1$:
\begin{eqnarray} \label{summary}
c_1=c_2=c_3 \nonumber
\end{eqnarray}

At this point we should make two observations. As claimed the
terms proportional to $c_4$ and $c_5$ remain undetermined. This is 
not obvious a priori although the proposed langrangian (\ref{LBachas}) 
requires the appearance of such terms. In the next section we will try
to give an explanation to the fact that $c_4$ and $c_5$ remain
undetermined. The second observation has to do with the other
three coefficients $c_1, c_2$ and $c_3$ which are equal to each
other. Using (\ref{summary}) in (\ref{Lambi}) and in view of the
Gauss-Codazzi equations (\ref{GaussCod}), they combine to form a
"Gauss-Bonnet" type lagrangian for the induced metric:
\begin{eqnarray} \label{LGBinduced}
&\displaystyle \frac{\sqrt{\tilde{g}}}{32 \pi ^2}L^{ambig}= \tilde{L}^{(T)}_{GB}= \nonumber \\
&\displaystyle \frac{\sqrt{\tilde{g}}}{32 \pi^2} \blp (R_T)_{\alpha \beta
\gamma \delta }(R_T)^{\alpha \beta \gamma \delta } -4
(\hat{R}_T)_{\alpha \beta }(\hat{R}_T)^{\alpha \beta }+
(\hat{R}_T)^2 \brp
\end{eqnarray}
where we use $\tilde{L}^{(T)}$ to remind ourselves, that this
expression holds up to terms proportional to the trace of the
second fundamental form, which should appear in $L^{(T)}_{GB}$ since equations
(\ref{GaussCod}) dictate such terms for $(\hat{R}_T)_{\alpha \beta
}$ and $(\hat{R}_T)$.

The unknown coefficient $c_1$ has already been determined in
\cite{BachasCurvature} using a duality argument. We will shortly
review this argument for completeness. Type IIA string theory
compactified on a K3 manifold is known to be dual to heterotic
string theory on $T^4$ \cite{9410167}. Assume now that we have a
wrapped D4-brane on the surface K3. It is known that D4 branes
form bound states with D0 branes. Actually D4 branes carry D0
charge \cite{Vafa}.

Now duality with heterotic string theory leads to the following
mass formula on the type-IIA side \cite{Waldram} for a bound state
of D4 and D0 branes:
\begin{eqnarray} \label{massformula}
m= T_{(4)} n_4 V_{K3} + T_{(0)} n_0
\end{eqnarray}
where $n_4$ and $n_0$ are the $D4$ and $D0$ charges respectively.
The DBI action is proportional to the volume of the surface the
brane wraps. This reproduces the first part of the mass formula
(\ref{massformula}). The second part should be reproduced by the
$R^2$ terms of the action. For a brane wrapping on a K3 surface
only the first term $R_{\al \beta \gamma \de} R^{\al \beta \gamma
\de}$ contributes. The second terms vanishes since K3 is Ricci
flat and the last two because the normal bundle is trivial. The Gauss-Bonnet
lagrangian is proportional to the $R_{\al \beta \gamma \de} R^{\al \beta \gamma
\de}$ term for a Ricci flat manifold. As mentioned before $L_{GB}$ is a topological
invariant, the Euler number, for 4d manifolds. For K3 it takes the value:
\begin{eqnarray}\label{R2K3}
\chi \cong \int_{K3} \; d^4 x L_{GB} = \frac{1}{32 \pi^2}
\int_{K3} \; d^4x \sqrt{\tilde{g}} R_{\al \beta \gamma \de}R^{\al \beta
\gamma \de}=24
\end{eqnarray}
Therefore using $L_{(p)}$ from (\ref{CurvBI}) in (\ref{massformula}) we get:
\begin{eqnarray} \label{c1=0}
m= T_{(4)} V_{K3} + (4 \pi^2 \apr)^2 T_{(4)} (c_1+1)= T_{(4)}
V_{K3} + T_{(0)} (c_1+1)
\end{eqnarray}
Comparison with (\ref{massformula}) shows that $c_1=0$. This
consequently sets the $c_2$ and $c_3$ ambiguities to zero as well. We have
verified in this way that the lagrangian (\ref{LBachas}) proposed
by \cite{BachasCurvature} is indeed correct to order $(\apr)^2$
up to terms proportional to the trace of the second
fundamental form.


\section{Conclusions}\label{conclusion}

In this paper, we studied the structure of $R^2$ terms in the
D-brane effective action extending the analysis of \cite{BachasCurvature} for the 
case of non-geodesic $(\Omega \neq 0)$ embeddings. We compared the results of our string amplitude
computations with the action proposed in ref.\cite{BachasCurvature} supplemented by an ambiguous 
Langrangian term (\ref{Lambi}). We found that the string amplitudes relate the ambiguous coefficients $c_1, \
c_2, \ c_3$ of (\ref{Lambi}) among each other giving us two equations for the three
unknowns. These equations can be used to determine two of the coefficients, 
say $c_2, \ c_3$, in terms of the third one, $c_1$. Nevertheless, the string amplitudes 
alone do not fix the complete form of the action, leaving the coefficients 
$(c_4,c_5)$ and $c_1$, as mentioned above, undetermined.
The conjectured type $IIA-Heterotic$ duality was used to determine the
value of $c_1$. The duality argument holds for branes wrapped on K3, a geometry with trivial 
normal bundle on the D-brane. Our results though are valid for arbitrary normal bundles. 
It would be interesting to understand the importance of our corrections for type 
$IIA-Heterotic$ duality in the case of a nontrivial normal bundle.

Therefore we conclude that the  lagrangian proposed in \cite{BachasCurvature} reproduces
the ${\cal O}(\apr^{\; 2})$ terms of our string scattering amplitudes.
The guess that $R^2, R \Omega^2$
and $\Omega^4$ terms can be written in terms of invariants
constructed by the world-volume curvature $R_T$ is indeed correct 
up to terms which vanish in the lowest
order equations of motion for the graviton and scalars. Such terms
do not contribute when we consider linear expansion in the fields
for the Riemann tensor and second fundamental form. Nevertheless
there is no reason a priori that they should not be present in
higher point amplitudes where we need to expand the Riemann and
second fundamental form tensors beyond the leading order. Nevertheless 
cancellation of the contact and exchange contributions of those terms 
makes it impossible to determine their coefficients in the D-brane lagrangian. 

As promised earlier we will attempt to explain why the coefficients 
$c_4$ and $c_5$ remain undetermined. The equivalence theorem (see \cite{zuber, haag})
states that the S-matrix elements are invariant under field redefinitions. In other 
words although field redefinitions might change some coefficients in the 
lagrangian, the scattering process does not depend on such terms.
This was exactly the case in \cite{theisen} \footnote{
We wish to thank Pascal Bain for pointing out \cite{theisen} to us}
where in the context of heterotic
string theory similar cancelations between contact and exchange contributions
made it impossible to deteremine through string amplitude computations $R^2$ corrections
that involve $R_{\mu \nu}$ or $R$. 
It was actually demonstrated in \cite{theisen}
that even loop computations cannot determine the full structure of the 
higher derivative terms. The same cancelations between contact and exchange
diagrams persist in genus 1 superstring amplitudes and render the fixing of such
lagrangian terms impossible. It was actually conjectured that these cancelations
will continue at higher loop amplitudes as well. 
    
Following the discussion of \cite{Vanhove}, if the on-shell action at higher order in 
some coupling constant g ($\apr^2$ in our case) contains terms proportional to the 
equations of motion obtained from the lower-order action, as in example:
\begin{eqnarray} \label{Vanhove1}
S[\phi]= \displaystyle S_0[\phi] + g \int dx \frac{\de S_0[\phi]}{\de \phi(x)} R(\phi(x), \partial
 \phi(x))
 \end{eqnarray}
these can be removed by the field redefinition:
\begin{eqnarray} \label{Vanhove2}
\phi \to \phi - g R(\phi, \partial \phi) \rightarrow S[\phi] \to S_0[\phi] + {\cal O}(g^2)
\end{eqnarray} 
resulting in an ambiguity for such langrangian terms.
In our case it is a field redefinition of the scalar field $X^{i}$ 
that generates an infinite set of derivative terms proportional 
to $Tr\Omega$. Under the redefinition: 
\begin{eqnarray} \label{Xredef}
X^{i} \to
X^{i} + \frac{1}{24} \frac{ (4 \pi^2 \alpha ^{'} ) ^{2} }{ 32 \pi ^{2}} 
(q \Omega^i_{\al \beta} R^{\al \beta} +  r \Omega^i_{\al \beta}
\Omega^{j \; \beta \gamma} \Omega^{j \; \al}_{\gamma}) 
\end{eqnarray}
the effective lagrangian changes:
\begin{eqnarray}\label{Lvar}
\displaystyle \de L^{(p)}= \frac{T_{(p)}e^{-\phi} \sqrt{\tilde{g}}}{2}  \frac{ (4
\pi^2 \apr ) ^{2} }{ 24 \cdot 32 \pi ^{2}}
(\Omega^{i \; \al}_{\al}) ( q \Omega^i_{\al \beta}R^{\al \beta} +  r \Omega^i_{\al \beta}
\Omega^{j \; \beta \gamma} \Omega^{j \; \al}_{\gamma})+  {\cal O}(\apr^4)
\end{eqnarray}
where the variation of the lagrangian came from the variation 
of $\sqrt{\tilde{g}}$ and we have used the formula:
\begin{eqnarray} \label{deltag}
 \frac{\de \sqrt{\tilde{g}}}{\de X^i}=
\frac{1}{2} \sqrt{\tilde{g}} \Omega^{i \; \al}_{\al}
\end{eqnarray}
The above redefinition generates a symmetry of the S-matrix under the 
shifts: $\de c_4= \frac{r}{2}$ and $\de c_5= \frac{q}{2}$. As in \cite{theisen}
we expect this phenomenon to continue to higher genus(loop) computations.
In other words pertubation theory cannot help us fix the form of the 
lagrangian completely. We will have to perform some non-perturbative or 
off-shell computation in order to determine the ambiguous terms.

The whole discussion above does not mean of course that terms subject to change under field redefinitions are 
unimportant or can be set to any value we choose. As demonstarted in \cite{theisen} 
under field redefinitions of the graviton the lagrangians we get correspond to physically inequivalent theories.
In their case, depending on the value of the arbitrary coefficients, the $R^2$ corrections can be of the form
$(R_{\mu \nu \rho \sigma})^2$ or a Gauss-Bonnet term or even $(C_{\mu \nu \rho \sigma})^2$, 
where $C_{\mu \nu \rho \sigma}$ the Weyl tensor. But each choice leads to a different spectrum 
with the GB case giving a spin-2 graviton, the $R^2$ case in addition to the spin-2 graviton a 
scalar and the $C^2$ case ghost-like fields. In our case although there is no obvious 
difference in the low energy spectrum for different choices of the coefficients $c_4$ and $c_5$, it is 
in principle plausible that such terms will affect calculations involving branes in curved
backgrounds. For example they could proove crucial for studying thermal YM theories 
employing the $AdS/CFT$ correspondence. The consistency of the computations in \cite{TaylorKiritsis}, which involve
D-brane probes approaching the horizon of black holes in $AdS$ space-time, depends on the form of the acceleration
terms on the D-brane world-volume action. It is therefore important to determine the complete structure 
of the higher-derivative terms. 

Another explanation\footnote{This argument was suggested to us by I.Antoniadis and P.Vanhove}
for the cancelation of contact and exchange diagrams is that since 
we expanded the second fundamental form to the the next order in $\apr$  then the equations
of motion for the scalars receive $\apr$ corrections and become $\partial^2 X^i= 
{\cal O}(\apr)$. Therefore one needs to modify
the propagator of the scalar fields. We used the scalar propagator to the 
zeroth order in $\apr$ in our computations. By adding exchange and contact pieces 
we actualy solved diagrammaticaly the equations of motion for the scalar fields. 
To this order the equations of motion are still $Tr \Omega^i=0$ and therefore
terms proportional to $Tr \Omega$ do not contribute to our amplitudes.

The results first presented in \cite{BachasCurvature}
were used for checking various dualities in cases where the second
fundamental form $\Omega$ was vanishing. 
Now, having determined the complete form of the action, it would be interesting
to extend this analysis to non-geodesic
($\Omega\neq 0$) manifolds.

\acknowledgments{I am grateful to Tomasz Taylor for numerous suggestions and 
his continuous guidance during this project. 
I also wish to thank Ignatios Antoniadis, Costas Bachas, Pascal Bain, Konstandinos Sfetsos, Pierre
Vanhove for fruitful discussions and Ecole Polytechnique where the last stage of this work 
took place. This work was supported in part by NSF Grant PHY-99-01057 and a Chateaubriand Fellowship.}

\renewcommand{\thesection}{A}
\setcounter{equation}{0}
\renewcommand{\theequation}{A.\arabic{equation}}
\appendix
\section{Integral formulas 1}\label{integrals}

The integrals we need to evaluate for the string amplitude
(\ref{3+1Amp}) are of the general form:
\begin{eqnarray} \label{Integ}
I = \int_{\cal{H}^{+}} d^2 \!z |1-z|^{a} |z|^{b} (z - \bar{z})^{c}
(z + \bar{z})^{d}
\end{eqnarray}
The region of integration $\cal{H}^{+} $ is the upper half complex
plane. This integral is convergent only when
\begin{eqnarray} \label{ConvReg}
a+b+c \leq -2    \nonumber \\
a+b+d \leq -2
\end{eqnarray}
These conditions are not satisfied for the integrals we need to
calculate in section \ref{1cl3o}. We will calculate the
integral in the convergent region and then analytically continue
to the physical region.

In order to perform this integration we  use the well known trick
of converting the integrand to two integrals of gaussian form,
using the formulas
\begin{eqnarray} \label{Trick}
|z|^{b}= \frac{1}{\Gamma(- \frac{b}{2})} \int_{0}^{\infty} du \,u^{-\frac{b}{2} - 1} e^{-u |z|^2} \\
|1-z|^{a}= \frac{1}{\Gamma(- \frac{a}{2})} \int_{0}^{\infty}
ds\,s^{-\frac{a}{2} - 1} e^{-s |1-z|^2} \nonumber
\end{eqnarray}
By writing the complex integral over $z=x+iy$, as two real
integrals over the x and y axis the two integrals decouple. We can
use known formulas for gaussian integrals to perform the y
integration
\begin{eqnarray} \label{yInteg}
I_{y}= \int^{\infty}_{0} dy \; y^c e^{-(s+u)y^2} =
\frac{\Gamma(\frac{1+c}{2})}{2(s+u)^{\frac{1+c}{2}}}
\end{eqnarray}

The integral over x can be found using the generating functional
\begin{eqnarray}
&F(\lambda)= \displaystyle \int_{-\infty}^{\infty} dx e^{-(s+u)x^2 + 2 \lambda x} =\nonumber \\
& = \sqrt{\frac{\pi}{s+u}} e^{\frac{\lambda^2}{s+u}}
\end{eqnarray}
The integral over x is then given by
\begin{eqnarray}
I_x= 2^d e^{-\frac{us}{s+u}}\int^{\infty}_{-\infty} dx \; x^d
e^{-(s+u)(x-\frac{s}{s+u})^2}= e^{-s} \frac{d \;}{d^d \! \lambda}
F(\lambda)|_{\lambda=s}
\end{eqnarray}
The expression above gives the integral over x as a rational
function of $ s$ and $ u$ with an exponential factor in front.
Although we do not have this result in a closed form, all the
integrals in our calculations involve $d=n, \ \ n \in Z $ which can
be easily done for each case separately. The necessary integrals
are:
$$
\label{xInteg} I_{x} = 2^{d} e^{- \frac{us}{s+u}}
\frac{\sqrt{\pi}}{(s+u)^{\frac{1}{2}}} \left \{
\begin{array}{cc}
1 & , d=0 \\
\frac{u}{(s+u)} &, d=1
\end{array} \right.
$$

Following the method introduced by \cite{Schwartz} for calculating
the four closed string amplitude on the sphere we can perform the
integrals over s, u by making the change of variables
$$
\begin{array}{cc}
w= \frac{u}{s+u}  & , 0 \leq w \leq 1 \\
v= \frac{su}{s+u} & , 0 \leq w \leq \infty
\end{array}
$$
The final result is :
\begin{eqnarray} \label{FinInteg}
I = (2 \imath)^{c} 2^d \,  \pi \frac{ \Gamma( 1+ d +
\frac{b+c}{2})\Gamma( 1+ \frac{a+c}{2})\Gamma( -1-
\frac{a+b+c}{2})\Gamma( \frac{1+c}{2})}{
\Gamma(-\frac{a}{2})\Gamma(-\frac{b}{2})\Gamma(2+c+d+
\frac{a+b}{2})}
\end{eqnarray}
where $ d= 0,1$

\renewcommand{\thesection}{B}
\setcounter{equation}{0}
\renewcommand{\theequation}{B.\arabic{equation}}
\section{Some geometry of submanifolds}
\label{submanifold}

We follow the analysis of the geometrical features of submanifolds
as described in \cite{BachasCurvature, Eisen, Koba}. To describe the
embedding of a D-brane in the ambient space we introduce
coordinates $ \sigma^{\alpha}, (\: \alpha= 0, \dots ,p )$. The
fields $ X^{\mu}(\sigma) \: \mu=0,\dots,9 $ describe the embedding
of the p-brane in the ten dimensional space-time. The two vector
fields $ \partial_{\alpha} X^{\mu} \: and \: \xi_{i}^{\mu}, \:
(i=p+1,\dots,9) $ define tangent and normal bundle frames
respectively and satisfy the relations
\begin{eqnarray} \label{DefSub}
\xi_{i}^{\mu}\xi_{j}^{\nu} G_{\mu \nu} = \delta_{i j} \quad and
\quad \xi_{i}^{\mu}\partial_{\alpha} X^{\nu} G_{\mu \nu}=0
\end{eqnarray}
where $ \delta_{ij} $ is the normal bundle metric. Using these
frames we pull-back tensors of the ambient space to the tangent
and normal bundle. The pull-back metric has the form
\begin{eqnarray} \label{PullBackMet}
\tilde{g}_{\alpha \beta}(\sigma)=\partial_{\alpha} X^{\mu}\partial_{\beta}
X^{\nu} G_{\mu \nu}
\end{eqnarray}
We raise and lower world-volume indices using this metric and
similarly, for normal bundle indices using $ \delta_{ij}$. There
are three independent covariant derivatives defined with respect
to: the ambient space connection $ \Gamma_{\nu \rho}^{\mu} $,
tangent bundle connection $(\Gamma_{T})_{\alpha \beta}^{\gamma}$
and the normal bundle connection $ (\omega)_{\alpha}^{i j}$
\begin{eqnarray}
(\omega)_{\alpha}^{i j}=\xi^{\mu,[i}(G_{\mu \nu} \partial_{\alpha}
+ G_{\mu \sigma} \Gamma_{\nu \rho}^{\mu} \partial_{\alpha}
X^{\rho}) \xi^{\nu,j]}
\end{eqnarray}
which is defined by requiring the normal frame to be covariantly
constant. The ambient space and tangent bundle connections are the
usual Christoffel symbols of the corresponding metrics.

Differentiating covariantly (\ref{PullBackMet}) with respect to
the world-volume coordinates $\sigma$ the left hand side vanishes,
since it is the covariant derivative of the pull back metric with
respect to the world-volume connection. The right hand side is the
projection to the tangent bundle of a tensor, which we define as
the second fundamental form
\begin{eqnarray} \label{SecFun}
\Omega_{\alpha \beta}^{\mu}=\Omega_{\beta \alpha}^{\mu}=
\partial_{\alpha}\partial_{\beta} X^{\mu} - (\Gamma_{T})_{\alpha
\beta}^{\gamma} \partial_{\gamma} X^{\mu} + \Gamma_{\nu
\rho}^{\mu} \partial_{\alpha}X^{\mu}
\partial_{\beta}X^{\nu}
\end{eqnarray}
This is a vector of the ambient space and a second rank
world-volume tensor. However  by construction the second
fundamental form has a vanishing projection to the tangent bundle
therefore we need only to consider the normal bundle projection
\begin{eqnarray} \label{NormProj}
\Omega_{\alpha \beta}^{i}= \Omega_{\alpha \beta}^{\mu}
\xi_{\mu}^{i}
\end{eqnarray}

There are two ways to construct Riemann tensors which transform
covariantly under world-volume reparametrizations and normal frame
rotations. One way is to pull-back the ambient space Riemann
tensors and the second by constructing Riemann tensors from
world-volume and normal bundle connections. However the
Gauss-Codazzi equations relate these two kinds of tensors
\begin{eqnarray} \label{GaussCod}
&(R_{T})_{\alpha \beta \gamma \delta}=R_{\alpha \beta \gamma
\delta} + \delta_{ij}(
\Omega_{\alpha \gamma}^{i} \Omega_{\beta \delta}^{j}-\Omega_{\alpha \delta}^{i} \Omega_{\beta \gamma}^{j}) \\
&and \nonumber \\
&(R_{N})_{\alpha \beta}^{\ \ \ ij}= R_{\alpha \beta}^{ \ \ \ ij} +
g^{\gamma \delta}( \Omega_{\alpha \gamma}^{i} \Omega_{\beta
\delta}^{j}-\Omega_{\alpha \gamma}^{j} \Omega_{\beta \delta}^{i})
\nonumber
\end{eqnarray}

 When writing possible ${\cal O}(\alpha^{'2})$
invariants we need to keep in mind that we cannot determine , by
comparison with string amplitudes, terms which vanish due to the
lowest-order equations of motion since the string amplitudes are
evaluated on-shell. Therefore, since the lowest order equations of
motion impose the vanishing of the ambient space Ricci tensor and
of the trace of the second fundamental form at linearized level,
\begin{eqnarray} \label{EQMotion}
R_{\mu \nu} = 0  \qquad and \qquad \Omega_{\; \alpha}^{\mu \,
\alpha}=0
\end{eqnarray}
invariants involving these tensors are non-vanishing only if we
expand them beyond the linear approximation. The $(curvature)^2$
terms determined in \cite{BachasCurvature} are modulo invariants
which involve these tensors since in their case all field theory
vertices were derived from the linearized form of the curvature
and second fundamental form tensors. Expanding around a flat
static space-time and using the static gauge for the embedding
\begin{eqnarray} \label{StaticGauge}
X^{\mu}(\sigma)=( \, \sigma^0,\dots
\sigma^p,X^{p+1}(\sigma),\dots,X^{p+1}(\sigma)\,)
\end{eqnarray}
equations (\ref{EQMotion}) imply the mass-shell conditions on
graviton and scalar field respectively. As a result of the first
condition in (\ref{EQMotion}) we only need to consider Ricci
tensors $\hat{R}_{\alpha \beta}$ $\hat{R}_{ij}$ constructed from
contractions of world-volume indices of the pull-back curvature
tensors. We use the hat on these tensors to distinguish them from
pull-backs of Ricci tensors of the ambient space.
 In addition the possible invariants
are constrained by the symmetries of the Riemann tensor and the
cyclic permutation property
\begin{eqnarray}
R^{\mu}_{[\nu \rho \sigma]}=0
\end{eqnarray}

As a result of these constrains not all pull-backs of the ambient
space Riemann tensor are independent. Eventually, it turns out
that only six pull-backs of the Riemann tensor
\begin{eqnarray}
&R_{\alpha \beta \gamma \delta}, \; R_{\alpha \beta \gamma i}, \; R_{\alpha \beta ij}, \; \nonumber \\
&R_{\alpha \{i j \} \beta}, \; R_{\alpha ijk}, \; R_{ijkl}, \;
\end{eqnarray}
and four contractions of these pull-backs
\begin{eqnarray}
\hat{R}_{\alpha \beta}, \; \hat{R}_{\alpha i}, \; \hat{R}_{ij}, \;
\hat{R}, \;
\end{eqnarray}
are independent. 

The possible  ${\cal O}(\alpha^{'2})$ invariants  need
to satisfy world-volume reparametrizations and normal frame
rotations invariance. Therefore we need to consider full
contractions of the tensors enumerated above, among themselves and
with the second fundamental form. There are ten $ R^2$, six $R
\Omega^2$ and four $\Omega^4$ invariants. The ten $R^2$ terms are
not completely independent with each other, that is, for linear
expansion of the $R's$ in the graviton field. There is one
ambiguity which is the Gauss-Bonnet term. It turns out that
similarly there are ambiguous combinations for the $ R \Omega^2$
and $ \Omega^4$ terms all reducing to total derivatives at
linearized level.  For completeness we list all the
$R^2$, $R \Omega^2$ and $ \Omega^4 $ terms: \\
squares of the pull-back Riemann  and $\hat{R}$ tensors
\begin{eqnarray}\label{R2}
&R_{\alpha\beta\gamma\delta} R^{\alpha\beta\gamma\delta}, \quad
R_{\alpha\beta\gamma i}R^{\alpha\beta\gamma i}, \quad
R_{\alpha\beta i j} R^{\alpha\beta i j}, \cr &R_{\alpha \{ ij\}
\beta} R^{\alpha \{ ij\} \beta}, \quad R_{\alpha ijk} R^{\alpha
ijk}, \quad  R_{ ijkl} R^{ ijkl}, \cr &{\hat R}_{\alpha\beta}
{\hat R}^{\alpha\beta}, \qquad
 {\hat R}_{\alpha i}{\hat R}^{\alpha i}, \qquad
{\hat R}_{i j}{\hat R}^{i j}, \qquad  {\hat R}^2.
\end{eqnarray}
and terms involving the second fundamental form
\begin{eqnarray} \label{ROmega2}
 &R^{\alpha\beta\gamma\delta}
(\Omega_{\alpha\gamma} \!\cdot \! \Omega_{\beta\delta}), \qquad
R^{\alpha\beta}_{\ \  i j}\; \Omega^i_{\alpha\gamma} \Omega^{j\
\gamma}_{\beta}, \qquad R^{\alpha\ \  \ \ \beta}_{\  \{i j\}}\;
 \Omega^i_{\alpha\gamma} \Omega^{j\ \gamma}_{\beta}, \cr
&{\hat R}^{\alpha\beta} (\Omega_{\alpha\gamma} \!\cdot \!
\Omega_{\beta}^{\ \gamma}),\qquad\quad {\hat R}_{i j}\;
\Omega^i_{\alpha\beta} \Omega^{j\ \alpha\beta}, \qquad\quad {\hat
R}\; (\Omega_{\alpha\beta} \!\cdot \! \Omega^{\alpha\beta}) .
\end{eqnarray}
 and
\begin{eqnarray} \label{Omega4}
&(\Omega_{\alpha\beta}\!\cdot \! \Omega^{\alpha\beta})
(\Omega_{\gamma\delta} \!\cdot \! \Omega^{\gamma\delta}), \qquad
(\Omega_{\alpha\gamma} \!\cdot \! \Omega^{\alpha\delta})
(\Omega^{\beta\gamma} \!\cdot \! \Omega_{\beta\delta}), \cr
&(\Omega_{\alpha\beta} \!\cdot \! \Omega_{\gamma\delta})
(\Omega^{\alpha\beta} \!\cdot \! \Omega^{\gamma\delta}), \qquad
(\Omega_{\alpha\beta} \!\cdot \! \Omega_{\gamma\delta})
(\Omega^{\alpha\gamma} \!\cdot \! \Omega^{\beta\delta}),
\end{eqnarray}

The ambiguous combinations which vanish to linear approximation to
the fields are
\begin{eqnarray} \label{TotalDer}
&R_{\alpha \beta \gamma \delta }R^{\alpha \beta \gamma \delta } -4
\hat{R}_{\alpha \beta
}\hat{R}^{\alpha \beta }+ \hat{R}^2 ,\nonumber \\
\nonumber \\
&4 R_{\alpha \beta \gamma \delta} (\Omega^{\alpha \gamma} \cdot
\Omega ^{\beta \delta}) -8 \hat{R}_{\alpha \beta} (\Omega^{\alpha
\gamma} \cdot \Omega ^{\;\beta}_{\gamma})+2 \hat{R}(\Omega^{\alpha
\beta} \cdot \Omega _{\alpha \beta})  \nonumber \\
&2(\Omega_{\alpha \gamma} \cdot \Omega_{\beta \delta})
(\Omega^{\alpha \gamma} \cdot \Omega^{\beta \delta}) -2
(\Omega_{\alpha \gamma} \cdot \Omega_{\beta \delta})
(\Omega^{\alpha \delta} \cdot \Omega^{\beta \gamma}),  \\
\nonumber \\
&-4(\Omega_{\alpha \gamma} \cdot \Omega_{\beta}^{\; \gamma})
(\Omega^{\alpha \gamma} \cdot \Omega^{\;\beta}_{\gamma}) +
(\Omega^{\alpha \beta} \cdot \Omega_{\alpha \beta})(\Omega^{\gamma
\delta} \cdot \Omega_{\gamma \delta}) \nonumber
\end{eqnarray}

In \cite{BachasCurvature} the various invariants were combined using the
Gauss-Codazzi equations, to quadratic invariants of $(R_T)$ and
$(R_N)$. We rewrite the expression involving $(R_T)_{\alpha
\beta}$, for the lagrangian of \cite{BachasCurvature}, in terms of the
invariants listed above. We do so, because it is more
convenient to identify the vertices, which
contribute to the field theory amplitudes, with the
lagrangian in this form. The amplitudes of section \ref{corrections1} and
\ref{corrections2} require that the second fundamental form
is expanded to the subleading order. In the static gauge this is
equivalent to a graviton field as it is shown in section
\ref{corrections1}. At this point we need to consider two
more invariants, which involve the trace of the second fundamental
form.
\begin{eqnarray} \label{Ambiguities}
(\Omega_{\gamma}^{\; \gamma} \cdot \Omega_{ \alpha}^{\;
\beta})(\Omega^{\; \alpha}_{\delta} \cdot \Omega_{\beta}^{\;
\delta})  \qquad \hat{R}_{\alpha \beta} (\Omega^{\alpha \beta}
\cdot \Omega ^{\; \gamma}_{\gamma})
\end{eqnarray}

 Finally, expanding  pull-backs of  Riemann tensors
we need the expressions of the tangent and normal bundle frames in
terms of open string modes (scalar fields). Although the tangent
bundle frame is written explicitly in terms of the fields in the
Born-Infeld action
\begin{eqnarray} \label{TangFrame}
\partial_{\alpha}X^{\mu}= \delta_{\alpha}^{\mu} + \delta_{i}^{\mu}
\partial_{\alpha} X_{i}
\end{eqnarray}
the normal bundle frame is not. We can use equations
(\ref{DefSub}) to solve perturbatively for $\xi_{i}^{\mu}$. We
assume that $\xi_{i}^{\mu}$ has convergent expansion in powers of
the fields $X^i$ which represent small fluctuations of the
position of the brane in the transverse space. We rescale the
fields $X^i \to \lambda X^i$, where $\lambda$ a small parameter.
Consequently, we write $\xi_{i}^{\mu}$ as an expansion in powers
of $\lambda$
\begin{eqnarray} \label{NormBundExp}
\xi_{i}^{\mu}= \xi_{i|0}^{\mu}+ \lambda \xi_{i|1}^{\mu} +
O(\lambda^2)
\end{eqnarray}
By plugging this expansion in to the second of (\ref{DefSub}) and
retaining terms up to $ \lambda $ we have
\begin{eqnarray} \label{SecEq}
\xi_{\alpha | 0}^{i} + \lambda( \partial_{\alpha}X^{j} \xi_{j |
0}^{i}+ \xi_{\alpha | 1}^{i}) + O( \lambda^2)=0
\end{eqnarray}
and the first of (\ref{DefSub}) implies
\begin{eqnarray} \label{SecEq2}
\xi_{i | 0}^{k} \xi_{j | 0}^{l} \delta_{kl} + O(\lambda^2)
=\delta_{ij}
\end{eqnarray}
Solving these two equations we get the following solution to
subleading order in $\lambda $
\begin{eqnarray} \label{SolExp}
\xi_{i}^{\mu}=\delta_{i}^{\mu} - \lambda \delta_{\alpha}^{\mu}
\partial^{\alpha} X_{i} + O(\lambda^2)
\end{eqnarray}

In our analysis, of the permissible invariants that contribute to
the graviton and three scalars amplitude, we have excluded terms
of the type $ D^2R, \: \Omega D^2 \Omega, \\
\Omega^2 D \Omega \ \
and \ \ \Omega DR $ which are corrections to the one-point
function of the graviton, scalar propagator and graviton-scalar
mixing. These vertices are presumably protected by supersymmetry
\cite{BachasCurvature}  and do not receive derivative corrections. It
can also be checked explicitly that such terms make unacceptable
contributions to the string amplitudes considered in \cite{BachasCurvature}
as well as to our amplitudes.

\renewcommand{\thesection}{C}
\setcounter{equation}{0}
\renewcommand{\theequation}{C.\arabic{equation}}
\section{Integral formulas 2}\label{integrals2}

The integrals we have to compute in section \ref{2cl1o} are
of the general form:
\begin{eqnarray}\label{generalintegral2}
I(a,b,c;P(z,\bz))=\int_{|z| <1} d^2 \! z |1-z|^{2a} |z|^{2b}
(1-|z|^2)^c \; P(z, \bz)
\end{eqnarray}
where $P(z,\bz)$ is a polynomial in $z, \bz$. Using polar
coordinates the expression above can be written:
\begin{eqnarray}\label{polarintegral}
\int_0^1 r \! dr \int_0^{2 \pi} d \theta (1-cos \theta + r^2)^a
r^{2b} (1-r^2)^c \; P(r cos \theta, r sin \theta)
\end{eqnarray}
We can compute the integrals over $\theta$ term by term for each
polynomial $P$ using formulas from \cite{Bateman}. Our
amplitude involves only polynomials up to first power in $rcos
\theta$ so we can use for the integral over $\theta$:
\begin{eqnarray}\label{thetaintegral}
&\displaystyle \int_0^{2 \pi} d \theta cos n \theta \; (1-2r cos \theta + r^2)^a
= \nonumber \\
&2 \pi \frac{\Gamma(n-a)}{\Gamma(-a)}  \frac{r^n}{n!}
F_{21}(-a,n-a;n+1;r^2)
\end{eqnarray}
where $F_{21}$ the hypergeometric function and $n=0,1$. Using
tabulated formulas from \cite{Grads} the integral over $r$
takes the general form:
\begin{eqnarray}\label{rintegral}
&\displaystyle \int_0^1 dr r^{2b+1+n} (1-r^2)^c \; F_{21}(-a, n-a;n+1;r^2)= \nonumber \\
&\frac{1}{2} \frac{\Gamma(1 + \frac{2b+n}{2})
\Gamma(1+c)}{\Gamma(2+c+\frac{2b+n}{2})}
F_{32}(-a,n-a,1+\frac{2b+n}{2};n+1,2+c+\frac{2b+n}{2};1)
\end{eqnarray}
where the generalized hypergeometric function $F_{32}$ is defined
in \cite{Bateman}. Combining the above results we get the
integral formula:
\begin{eqnarray}\label{formula}
&I(a,b,c; (r cos \theta)^n)=  \\
&\displaystyle \frac{\pi}{n!} \frac{\Gamma(n-a)}{\Gamma(-a)} \frac{\Gamma(1 +
\frac{2b+n}{2}) \Gamma(1+c)}{\Gamma(2+c+\frac{2b+n}{2})}
F_{32}(-a,n-a,1+\frac{2b+n}{2};n+1,2+c+\frac{2b+n}{2};1)\nonumber
\end{eqnarray}
Define the following integrals:
\begin{eqnarray}\label{integralsfinal}
&I_1=I(-s-t-u-1,t-1,u;z+\bz)=2I(-s-t-u-1,t-1,u;r cos \theta)
\nonumber \\
&I_2=I(-s-t-u-1,t-1,u+1;1) \nonumber \\
&I_3=I(-s-t-u,t-1,u;1) \\
&I(-s-t-u-1,t-1,u;(1+z)(1+\bz))= 2 I_1 + I_3 \nonumber
\end{eqnarray}
The integrals of (\ref{integrandfinal}) can be evaluated using the
expressions above.


\end{document}